\documentclass[10pt]{article}

\usepackage[a4paper,margin=1.9cm]{geometry}

\usepackage{float}
\usepackage{graphicx}%
\usepackage{subcaption}
\usepackage{multirow}%
\usepackage{amsmath,amssymb,amsfonts}%
\usepackage{amsthm}%
\usepackage{mathrsfs}%
\usepackage[title]{appendix}%
\usepackage{xcolor}%
\usepackage{textcomp}%
\usepackage{manyfoot}%
\usepackage{booktabs}%
\usepackage{algorithm}%
\usepackage{algorithmicx}%
\usepackage{algpseudocode}%
\usepackage{listings}%
\usepackage{subfiles}
\usepackage{placeins}
\usepackage[hidelinks]{hyperref}
\usepackage{natbib}
\usepackage{adjustbox}

\newcommand{\secref}[1]{\hyperref[#1]{\nameref*{#1}}} 

\graphicspath{{Fig/}}

\theoremstyle{thmstyleone}%
\theoremstyle{thmstyletwo}%
\theoremstyle{thmstylethree}%

\usepackage{kotex} 
\usepackage{fontspec}
\usepackage{xunicode}
\usepackage{xltxtra}

\usepackage{bbm} 

\usepackage{polyglossia}
\setotherlanguages{thai} 
\newfontfamily{\thaifont}{Garuda}[Extension=.ttf, Script=Thai]

\usepackage{titlesec}

\titleformat{\section}
  {\normalfont\bfseries\large}
  {\thesection.}
  {0.5em}
  {}

\titleformat{\subsection}
  {\normalfont\bfseries\small}
  {\thesubsection.}
  {0.5em}
  {}

\titleformat{\subsubsection}
  {\normalfont\itshape\small}
  {\thesubsubsection.}
  {0.5em}
  {}

\usepackage{authblk}       

\graphicspath{{Fig/}}

\author[1,2,$\ast$]{Bridget Smart}
\author[1]{Renaud Lambiotte}
\author[3]{Takaaki Aoki}
\author[4]{Ryota Kobayashi}

\affil[1]{\textit{Mathematical Institute, University of Oxford, Woodstock Road, Oxford, United Kingdom}}
\affil[2]{\textit{Institute for New Economic Thinking, University of Oxford, Manor Road, Oxford, United Kingdom}}
\affil[3]{\textit{Faculty of Data Science, Shiga University, Shiga, Japan}}
\affil[4]{\textit{Graduate School of Frontier Sciences, University of Tokyo, Tokyo, Japan}}
\affil[$\ast$]{\texttt{bridget.smart@maths.ox.ac.uk}}

\title{Inferring Temporal Dependencies from Social Time Series with the Cross-Correlogram}

\date{} 

\begin{document}

    \maketitle
    \begin{abstract}
      Characterizing temporal interactions in social systems is challenging because social behavior can be bursty and non-stationary, violating the stationarity assumptions of many methods used to measure temporal dependence. The cross-correlogram, an existing technique used to profile neural excitations and inhibitions, offers an interpretable alternative to methods such as Granger causality or co-occurrence, as it produces a full profile of lagged dependence directly from event times rather than a single summary statistic. We adapt the cross-correlogram by integrating functional models of behavior with data-driven temporal response profiling. By characterizing how periodic structure biases traditional cross-correlograms, we propose a correction based on smooth intensity functions, specified from a known functional form or estimated empirically. This approach provides a robust, interpretable estimator of temporal dependency profiles even when collective rhythms operate on timescales that overlap those of the interactions of interest. We demonstrate theoretically and through simulation that the proposed method recovers temporal dependencies in periodic regimes, outperforming interval-jitter and Granger causality methods. Finally, we apply the method to 3.1 million event times from X (formerly Twitter) collected between 2019 and 2020, demonstrating how cross-correlograms reveal delayed temporal relationships in collective online behavior that are missed by co-occurrence measures. For a subset of television-related hashtags, recovered delays align with known broadcast schedules, providing evidence that the proposed method captures genuine temporal structure rather than artifacts of shared attention cycles.
    \end{abstract}

    \vspace{1ex}


    \vspace{4ex}


\renewcommand{\thefootnote}{\arabic{footnote}}

\section{Introduction}
Inferring pairwise relationships from event time series is a fundamental task for quantifying interactions and uncovering patterns of dependence. This problem arises across many disciplines, including in neuroscience to study neural interactions \citep{perkel1967neuronal,tuckwell1988introduction,kobayashiInferenceMonosynapticConnections2024}, in finance to analyze market dependencies \citep{wangForecastingChinaCrude2021,bauwensModellingFinancialHigh2009}, and in geoscience to predict natural disasters \citep{reinhartReviewSelfExcitingSpatioTemporal2018,ogataSpaceTimePointProcessModels1998}. Across these domains, association measures are interpreted within well-specified mechanistic models, providing a basis for plausible causal inference. 
In contrast, system-level mechanisms in the social sciences are often under specified or poorly understood \citep{hedstromCausalMechanismsSocial2010,wattsCommonSenseSociological2014}. As a result, temporal measures of association tend to emphasize coordination or co-occurrence rather than revealing underlying generative processes . Although strong theoretical foundations exist for regular behavioral patterns, such as circadian, weekly, and seasonal rhythms \citep{aledavoodChannelSpecificDailyPatterns2016}, such regularities are rarely incorporated into system-level analyses. This omission has important consequences as association-based methods may conflate correlations arising from shared exogenous factors, such as collective attention cycles, with evidence of direct causal interaction \citep{moralesGlobalPatternsSynchronization2017}. This motivates a central question for temporal inference in social systems: how can we distinguish genuine directional relationships from correlations driven by shared rhythms or external influences?

We address this question by introducing the cross-correlogram as a general tool for recovering lagged dependencies in social systems. Although cross-correlograms were originally developed for experimental, multi-trial neural data \citep{kobayashiInferenceMonosynapticConnections2024}, their standard formulations rely on assumptions such as homogeneous baselines, time-scale separability, and stationarity that rarely hold in social settings. Characteristic features of social systems, including periodicity, burstiness, and broader non-stationarity can invalidate these assumptions, distort baseline structure, introduce spurious dependencies, and compromise standard significance tests.

By adapting the cross-correlogram to combine data-driven estimation with domain-specific behavioral mechanisms, we obtain full temporal profiles of dependence. We first quantify how periodic non-stationarity and burstiness systematically bias cross-correlogram estimates, then use these insights to propose corrections and outline a general procedure adaptable to any well-specified temporal pattern. 

Many existing corrections for temporal heterogeneity act on the data. Interval jitter, a widely used existing method which removes temporal structure at specified scales assumes linear, time-invariant processes and fails when the timescales of true dependencies overlap with those of unwanted temporal structure. We derive the error which a jittering procedure introduces into the cross-correlogram in periodic and bursty settings in \autoref{app:jittererror}

In this work we propose a correction that acts directly on the null distribution. The smooth intensity null models event rates as smooth functions within an inhomogeneous Poisson framework, specified either from a known functional form or estimated empirically. Event times are unmodified so the observed cross-correlogram retains every lagged dependency present in the data, with interactions accounted for in the expected coincidence counts. 


Our procedure complements existing co-occurrence measures in computational social science, which classify interactions within predefined time windows \citep{grahamVirusCoordinatedSpread2020,kellerPoliticalAstroturfingTwitter2020}, as well as directed measures such as Granger causality and transfer entropy, which infer predictive relationships \citep{grangerInvestigatingCausalRelations1969,shojaieGrangerCausalityReview2022,shortenEstimatingTransferEntropy2021}. In contrast, the cross-correlogram provides a symmetric, interpolation-free estimator that yields an interpretable lag distribution rather than a single summary statistic.

The remainder of the article is organized as follows. \secref{sec:relwork} reviews existing temporal approaches in neuroscience and computational social science. \secref{sec:framework} formalizes the cross-correlogram, discusses the challenge posed by non-stationarity and quantifies the effect of co-movement between periodic intensities. It then introduces the jitter and smooth intensity null corrections. We evaluate the functional and smooth intensity nulls on synthetic data with periodic non-stationarity in \secref{sec:validation} and assess robustness to bursty dynamics in \autoref{app:bursty}, comparing them against interval jitter, Granger causality and homogeneous baseline procedures. Alongside this method comparison, we test the impact of misspecified null models, providing evidence that a null model which under-represents the structure content of the true intensity (e.g. a unimodal periodic fit to bimodal circadian activity) produces high false-positive rates around 86\%, whereas a data-driven empirical correction remains valid without requiring a specified functional form. In \secref{sec:appdata}, we apply our procedure to 3.1 million posts from X (formerly Twitter), available at \cite{AUDATAMitchell2026}, constructing a network of delayed temporal relationships and validate the estimated response profiles against known broadcast delays. We also provide an open-source implementation at \url{https://github.com/bridget-smart/temporal_profiles}.

\section{Related work}\label{sec:relwork}

Temporal information is widely used across disciplines to measure and model relationships between processes. Correlations between event times can indicate physical dependencies \citep{reinhartReviewSelfExcitingSpatioTemporal2018,kobayashiReconstructingNeuronalCircuitry2019} or shared responses to external stimuli \citep{bauwensModellingFinancialHigh2009,parkInvestigatingClusteringViolence2021}. In computational social science, event times are used to infer \textit{coordination} between social media accounts, signaling automated or orchestrated activity \citep{pacheco2021uncovering,kellerPoliticalAstroturfingTwitter2020,gigliettoItTakesVillage2020}. Similar temporal data support clustering and prediction tasks such as civil unrest forecasting and hashtag analysis \citep{alsaediCanWePredict2017,wangTopicSentimentAnalysis2011}. In these applications, many methods target simultaneous or short-lag effects rather than profiling delayed temporal dependencies.

Classical tools such as cross-correlation, cointegration, Granger causality, and transfer entropy capture directional dependencies in continuous time series. Extensions of Granger causality \citep{shojaieGrangerCausalityReview2022} including those based on conditional intensities \citep{kim2011granger} and transfer entropy \citep{kobayashi2013impact,shortenEstimatingTransferEntropy2021} have been developed for irregular or discrete events; however, these methods typically rely on temporal binning or continuous approximations and require large sample sizes for reliable estimation. Each approach characterizes relationships along a distinct dimension: cross-correlation captures contemporaneous association, Granger causality quantifies linear predictive influence, transfer entropy measures nonlinear information transfer, and cointegration \citep{engle1987co} detects long-run equilibrium relationships. In contrast, the cross-correlogram framework presented here offers an intuitive, event-based representation of temporal dependencies in discrete processes. It captures asymmetric and potentially nonlinear lag structures directly from inter-event intervals, providing a complementary perspective to continuous or parametric approaches.

The cross-correlogram originates in neurophysiology, where it was introduced to characterize dependencies between simultaneously recorded spike trains \citep{perkelNeuronalSpikeTrains1967b}. Non-stationarity has long been a challenge, with lagged relations between processes producing structures in the cross-correlogram which may be incorrectly detected as causal relations. This has motivated corrections such as shuffle- and shift-predictors \citep{perkelNeuronalSpikeTrains1967b} and the joint peri-stimulus time histogram \citep{aertsen1989dynamics}, which estimate the component of the correlogram attributable to shared experimental conditions rather than to interaction. 
Brody \citep{brodyCorrelationsSynchrony1999} shows that covariation in firing rate or response latency may produce correlogram peaks which survive this correction. These methods all require repeated trials, and so are not appropriate for single-trial data, which is common in social settings.

A widely used alternative is jittering, a resampling procedure which removes temporal structure up to a prespecified timescale \citep{harrison2009rate}. Amarasingham et al.  \citep{amarasinghamConditionalModelingJitter2012} show that jittering corresponds to a conditional inference in which the rate is treated as approximately constant within each jitter window, while Stark and Abeles \citep{stark2009unbiased} show that the resulting null can absorb structure at the timescale of interest. We quantify this for periodic and bursty rate modulation in \autoref{app:jittererror}

Alongside effects due to periodicity and other co-movement, social activity is also characteristically bursty \citep{barabasi2005origin,karsai2018bursty}, and is modulated by circadian, weekly and seasonal rhythms \citep{aledavoodChannelSpecificDailyPatterns2016}. Both features have been incorporated into generative models of online activity, for example time-dependent Hawkes processes that combine an excitatory response with explicit circadian modulation to predict retweet dynamics \citep{kobayashiTiDeHTimeDependentHawkes2016a}. 

Related work extends pairwise temporal association measures to infer network structure from spike-train data using generalized linear models \citep{kobayashiReconstructingNeuronalCircuitry2019,truccolo2005point}, convolutional neural networks \citep{endoConvolutionalNeuralNetwork2021}, and community detection frameworks \citep{hoffmannCommunityDetectionNetworks2020}. More broadly, parallels between neural firing and hashtag timing have been noted \citep{sanliLocalVariationHashtag2015}, and Hawkes processes have been used to detect coordinated communities \citep{arastuieCHIPHawkesProcess2020}. Studies of hashtag diffusion and competition \citep{wengCompetitionMemesWorld2012,chenScalingLawsDynamics2020} have highlighted the richness of temporal dynamics in social data, though they focus on emergence and popularity rather than direct temporal interaction.

\section{Measuring Temporal Dependencies in Social Systems}
\label{sec:framework}

\begin{figure}[t]
    \centering
    \includegraphics[width=0.8\linewidth]{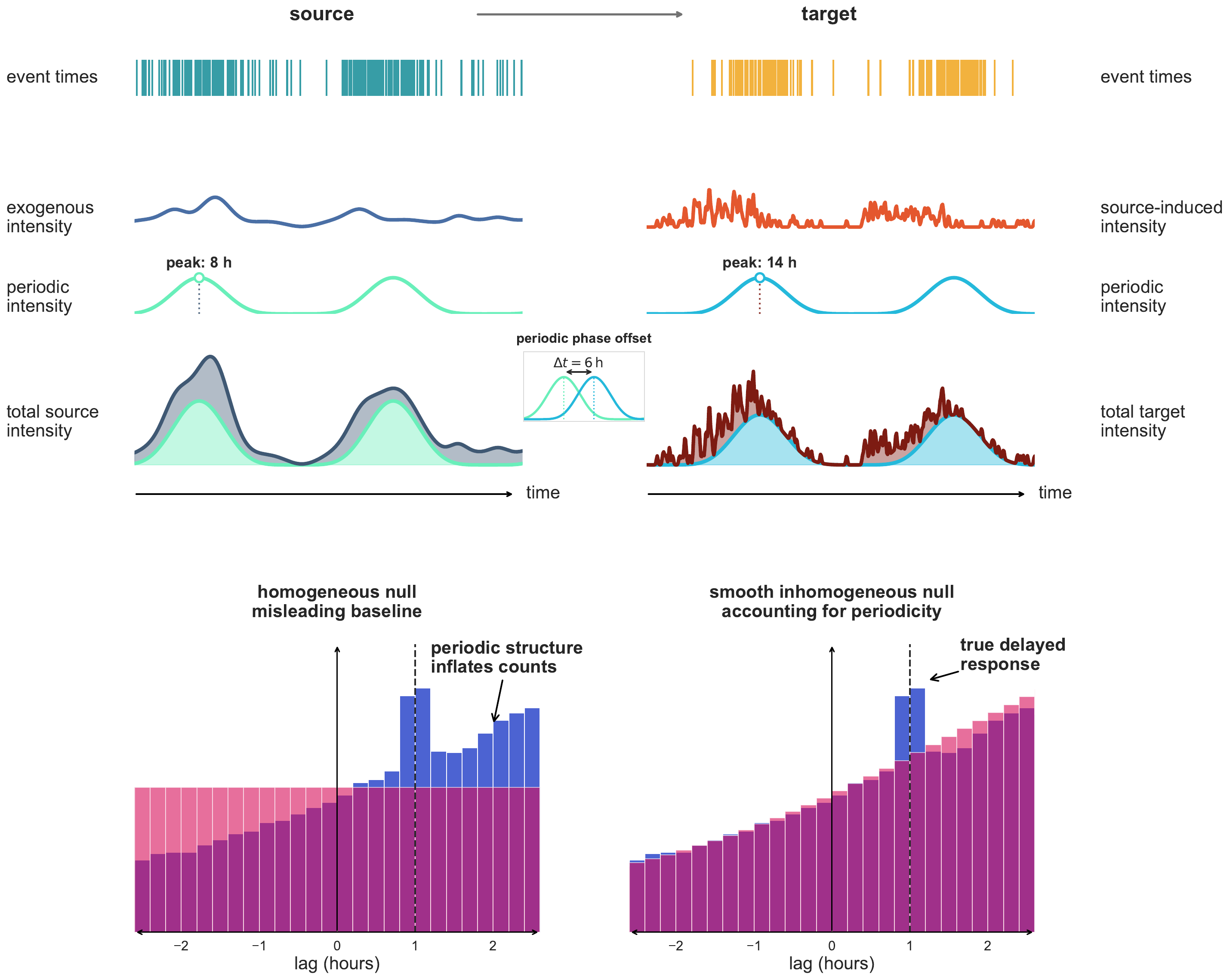}
    \caption{Simulated source and target point processes demonstrate how incorporating temporal structure into the null model separates true interaction from correlations induced by shared temporal variation. Simulated event times (top row) for the source (left) and target (right). The source intensity combines exogenous (dark blue) and periodic (cyan) components, while the target combines a periodic baseline (cyan) with a reactive component (dark orange) generated from delayed source events (delay $\sim \mathcal{N}(1,0.1^2)$). Total intensities are shown beneath the component decompositions. The bottom row compares cross-correlograms under homogeneous Poisson (left) and inhomogeneous (right) null models. Unlike the homogeneous null, the inhomogeneous null accounts for time-varying structure, isolating the induced interaction as a positive-lag peak near 1 hour. Bin heights are averaged over 30 simulations.}
    \label{fig:intuition}
\end{figure}

In this work we consider settings where events in a source process $X$ may modulate the instantaneous rate of a target process $Y$. Let $\{t_{x_1}, t_{x_2}, \ldots, t_{x_{N_X}}\}$ denote the event times of a source process $X$, and $\{t_{y_1}, t_{y_2}, \ldots, t_{y_{N_Y}}\}$ those of a target process $Y$. These temporal point processes are analogous to \textit{spike trains} in neuroscience.

Let $\lambda_Y(t)$ denote the conditional intensity of $Y$ at time $t$. We model the intensity function as
\begin{equation}\label{eq:intensity_form}
    \lambda_Y(t) = \lambda_Y^0(t) + \sum_{j=1}^{N_X} f(t - t_{x_j}),
\end{equation}
where $\lambda_Y^0(t)$ is the baseline rate of $Y$ and $f(\tau)$ is a response function describing how an event in $X$ affects the rate of $Y$ after a lag $\tau$, where $\tau= t - t_{x_j}$. Estimating $f(\tau)$ corresponds to inferring the temporal profile of dependence between the two processes. The cross-correlogram enables this profile to be estimated in the discrete setting.

Interactions between processes can arise from external stimuli, reactive mechanisms, or higher-order network dynamics. In physical and biological systems, well-characterized constraints often help identify relevant interaction timescales. In social systems, by contrast, actors are influenced by diverse internal and external factors, and the mechanisms driving behavior are rarely known. This complexity produces signals that are noisy, overlapping, and highly variable across time. The goal of this work is to adapt the cross-correlogram method to isolate temporal response profiles from these complex systems.

\subsection{\bf The cross-correlogram}

For every pair of events $x_j \in X$ and $y_k \in Y$, define the inter-event lag to be
\[
\delta_{k,j}^{(Y,X)} = t_{y_k} - t_{x_j}.
\]

The \textit{cross-correlogram} (CCH) is a histogram of these inter-event lags, approximating the lagged temporal relationship between the two processes. To avoid edge effects, we restrict source events to the interval $[w, T-w]$ and compute inter-event lags against all target events in $[0,T]$. Each bin $i$ has width $\Delta$ with edges at $b_i = -w + i\Delta$ for $i = 1, \ldots, N_b$, where $N_b = 2w / \Delta$. The height of bin $[b_i, b_i + \Delta)$ is
\begin{equation}
    h_{i}
   = \sum_{j=1}^{N_X} \sum_{k=1}^{N_Y}
      \mathbbm{1} \!\left(
         \delta_{k,j}^{(Y,X)} \in [b_i ,\, b_i + \Delta)
      \right),
\end{equation}
where $\mathbbm{1}(\cdot)$ is the indicator function. A peak at positive lag ($b_i > 0$) indicates that events in $X$ tend to precede those in $Y$ by approximately $b_i$ time units; a peak at negative lag suggests the reverse ordering. The window $[-w, w]$ is a hyperparameter chosen to match the timescale of the target behavior. 

Poisson processes are frequently used to model event times. In the homogeneous case, inter-event times are independent and identically distributed, making the process memoryless. For independent Poisson processes with sufficiently smooth intensity functions $\lambda_X(t)$ and $\lambda_Y(t)$, the expected cross-correlogram height for bin $i$ is
\begin{equation}\label{eq:smooth_int}
\mathbb{E}[h_{i}] = \int_w^{T-w} \int_{s+b_i}^{s+b_i + \Delta} \lambda_X(s)\,\lambda_Y(t)\, dt\, ds.
\end{equation}
A full derivation is provided in \autoref{app:binheights}.

When both processes are independent homogeneous Poisson processes with constant rates $\lambda_X$ and $\lambda_Y$, the contribution to each bin $i$ from each event in $X$ follows a $\text{Poisson}(\lambda_Y \Delta)$ distribution. Summing over all expected events in $X$, where $N_X \approx \lambda_X(T - 2w)$, we find the distribution of bin heights,
\[
h_{i} \sim \text{Poisson}\!\left(\lambda_X(T - 2w)\,\lambda_Y \Delta\right),
\]
and so
\[
\mathbb{E}[h_i] = \mathrm{Var}(h_i) 
   = \lambda_X(T - 2w)\,\lambda_Y \Delta.
\]
Under a Poisson distribution, the expectation and variance of the null distribution are equal and can be easily calculated. This fact is used to test the significance of observed coincidence counts.

Significance is typically assessed under an independent Poisson null \citep{shaoMeasureStatisticalTest1996}, using either per-bin tests or global goodness-of-fit statistics. Directional relationships are inferred from asymmetries in these response profiles; accordingly, only positive lags are evaluated when testing influence between a source and target process.

The homogeneous Poisson assumption, while mathematically convenient, is frequently violated in empirical systems. Event rates often exhibit strong non-stationarity from deterministic rhythms or bursty dynamics \citep{amarasinghamSpikeCountReliability2006, brodyArtefactualSpikeTrain1999}. In such settings, apparent correlogram peaks may reflect shared rate modulation rather than genuine interaction.

One way to address this is to perform significance testing against null models that explicitly permit specific classes of non-stationarity, including circadian, smooth rate-modulated, and jitter-based nulls. Each constrains the form of admissible temporal structure while preserving independence between processes. Under each null, parametric or empirical methods can be used to estimate local or global test statistics.

\subsection{\bf Non-Stationarity}\label{sec:nonstationarity}

\begin{figure}[t]
    \centering
    \includegraphics[width=0.6\linewidth]{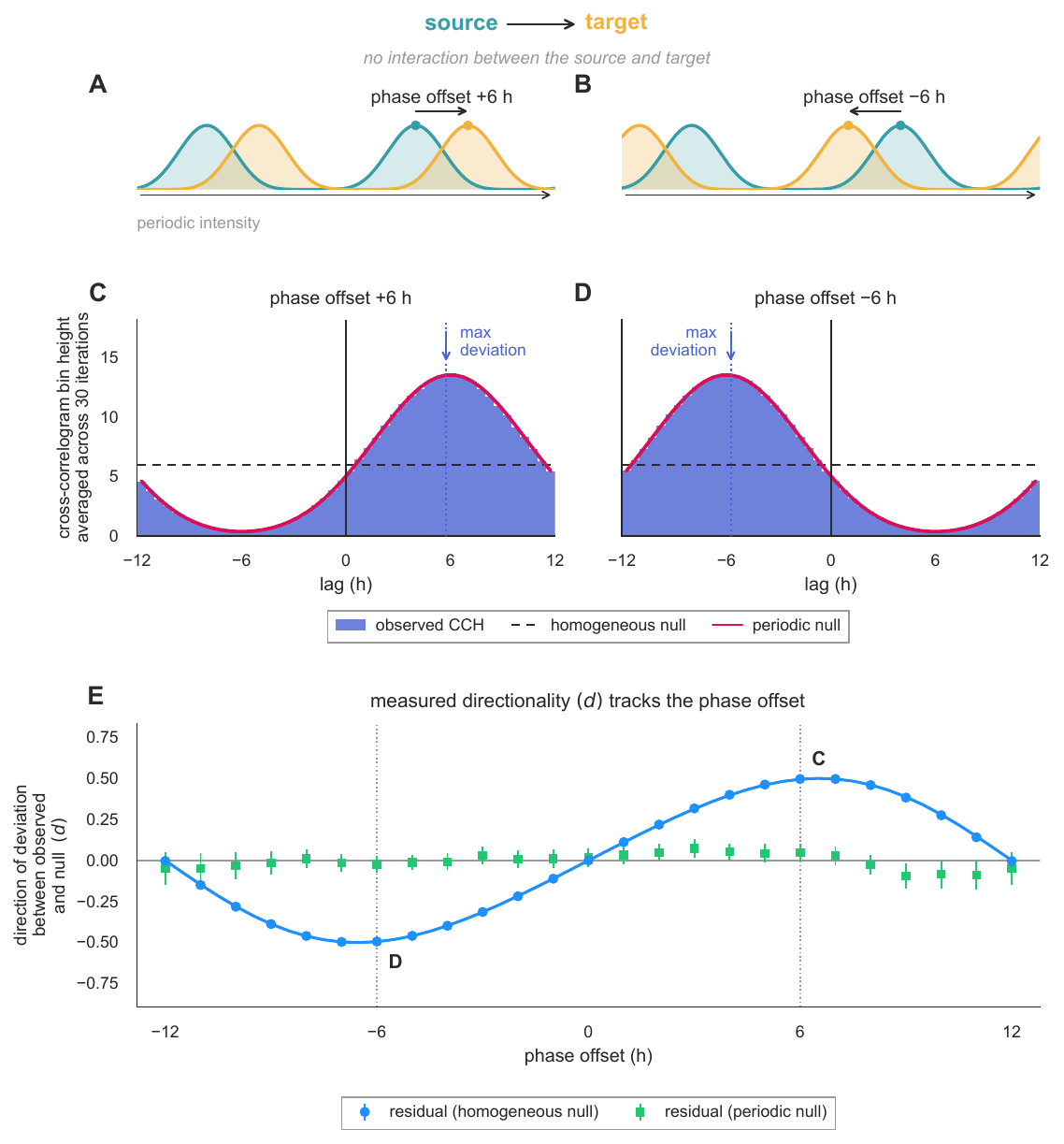}
    \caption{Co-movement produces directional structure in the cross-correlogram, shown for independent source and target processes with a $24$ hour periodic intensity. (A, B) Periodic intensities with offsets of $+6$ and $-6$ hours and the corresponding average cross-correlograms (C, D) ($30$ iterations, $T=504$). Even without a reactive component, each cross-correlogram shows a peak at the corresponding phase offset. (E) The normalized lag-weighted mean of the deviation between observed and expected bin counts in the cross-correlograms as a function of phase offset ($20$ iterations, error bars show standard error). Measured against the homogeneous null, directionality matches the phase offset. The effect is most pronounced at a phase shift of $90^\circ$ and zero at $0^\circ$ or $180^\circ$. A periodic null accounts for this structure and residuals are close to zero. Cross-correlograms are calculated using a $\pm 12$ hour window with $0.5$ hour bins, with mean rates of $1.5$ and $6$ events per hour for source and target respectively.}
    \label{fig:comovement}
\end{figure}

When event rates vary on timescales comparable to those of the interactions of interest, peaks in the cross-correlogram can arise from shared external rhythms rather than genuine
influence. Standard techniques for measuring correlations between temporal point processes assume that unwanted effects occur on scales distinct from the target behavior, that processes are linear and time-invariant, and that responses to stimuli remain constant. These assumptions rarely hold for social data, which span long timescales, generating mechanisms shift, and display non-stationary effects at scales overlapping with the behaviors of interest.

To motivate this work, we focus on two features prevalent in online social systems, periodic non-stationarity and burstiness. Periodic non-stationarity includes predictable cycles such as circadian, weekly or seasonal rhythms. Bursty behavior is characterized by clusters of frequent events separated by periods of low or no activity. Both phenomena operate on timescales comparable to social interactions, and therefore can produce cross-correlogram peaks that mimic true causal dependencies.

When the intensity functions of two processes exhibit periodic structure, lags between the peaks can result in peaks and troughs in the cross-correlogram which do not represent genuine interaction. The effects of this co-movement depend on the strength of the underlying periodic intensity, the pairwise offset and if the functions are in phase. \autoref{fig:comovement} illustrates this for a pair of independent inhomogeneous Poisson processes which share a $24$ hour periodic intensity which differs only in phase. There is no interaction between the processes but without correction, the observed cross-correlogram has a peak at the phase offset (\autoref{fig:comovement}C, D, E). 

These effects also interact with the corrections intended to remove them. Even under a constant Poisson null, jitter and smooth intensity corrections introduce biases of their own
which weaken statistical power and reliability. We quantify these biases through theoretical
error calculations in \autoref{sec:jitter} and \autoref{sec:SIF}, and through simulation in
\autoref{sec:validation} and \autoref{app:bursty}.

\subsection{\bf Interval Jitter}\label{sec:jitter}

Jitter is a procedure for removing unwanted temporal effects from cross-correlograms \cite{harrison2009rate,platkiewiczSpikeCenteredJitterCan2017,neimanIdentifyingTemporalCodes2011}. Its efficacy depends on the separation between the timescales of unwanted and target effects. Even in simplified simulations of bursty settings, we observe a strong relationship between the degree of burstiness, the amplitude of periodic non-stationarity, and the error introduced by jittering (\autoref{app:jittererror}).

In practice, social time series exhibit multiple sources of temporal structure simultaneously, and these effects can combine in complex ways. With compounding error and bias from both predictable and stochastic temporal features, the additional error introduced by jittering can substantially degrade cross-correlogram quality.

Applying an interval-jitter transformation is equivalent to smoothing the underlying intensity function by averaging over intervals of length $\delta_j>0$ \citep{amarasinghamConditionalModelingJitter2012}, treating events as equally likely at any point within each window. As the jitter interval approaches zero ($\delta_j \to 0$), the original intensity is recovered; as it approaches the observation period ($\delta_j \to T$), the transformation converges to a time-homogeneous approximation.

Let $\tilde{\lambda}_Y(t)$ denote the jittered intensity function. For any $t \in [0,T]$, it can be expressed in terms of the original intensity as
\begin{equation}
\tilde{\lambda}_Y(t) = \dfrac{1}{\delta} \int_{\delta \lfloor t/\delta\rfloor}^{\delta (\lfloor t/\delta\rfloor+1)} \lambda_Y(u)\, du.
\end{equation}

The theoretical error introduced by jittering in both non-stationary and bursty regimes is derived and validated using synthetic data in \autoref{app:jittererror}.

\subsection{\bf Smooth intensity function}\label{sec:SIF}

Unlike interval-jitter techniques that modify event times directly, we propose a method that incorporates temporal structure into the null, accounting for this structure in the expected bin heights. The observed cross-correlogram therefore retains all lagged dependencies, and only the significance threshold changes. We compute the expected bin heights using continuous-time approximations of the intensity function to model the effects of predictable or stochastic temporal structure on the null distribution, rather than removing them from the data. This approach follows \citep{parkEfficientAlgorithmContinuous2008}, who use continuous-time approximations of the cross-correlogram to locate events more precisely. 

For a sufficiently smooth null intensity, the theoretical bin heights of the cross-correlogram can be derived analytically (\autoref{eq:smooth_int}). This allows the null model to incorporate various representations of temporal structure. Within this approach we consider two complementary formulations: a \textit{functional} intensity model, in which external effects follow a known functional form, and an \textit{empirical} intensity model, which estimates rate variation directly from the data.

By comparing observed cross-correlograms against a theoretically informed inhomogeneous Poisson null, this framework provides a more realistic and interpretable null distribution than the homogeneous alternative, accommodating temporal structure arising from both predictable and stochastic processes. 

When domain knowledge provides a well-specified functional form for external effects, these can be incorporated directly into the smooth-intensity null. Many forecasting and signal-processing methods already account for regular trends such as circadian, weekly, or seasonal cycles.

When functional forms are unavailable, data-driven approaches offer a flexible alternative. Sliding window or interval averaging can be used to construct smoothed intensity functions that filter out fluctuations at specified temporal scales. When the intensity of the inhomogeneous Poisson null is estimated using an interval-based moving average, the result is equivalent to applying an interval-jitter transformation, but with the adjustment incorporated into the null model rather than the data.

To illustrate how temporal structure affects cross-correlogram estimation, we simulate a pair of point processes with underlying but offset periodic intensities. The offset between these intensities leads to artificial structure in the observed cross-correlogram, which can obscure the temporal response relationship between the processes. The source process combines exogenous and periodic components, while the target process includes a periodic baseline together with a `reactive' spike, corresponding to a subset of source events with a delayed response.

Figure~\ref{fig:intuition} shows the resulting event times, intensity decompositions, and cross-correlograms. When evaluated under a homogeneous Poisson null, the cross-correlogram fails to account for temporal variation in the underlying intensities, producing a flat baseline that obscures the true reactive relationship between the processes. In contrast, when exogenous and periodic structures are accounted for, the true response relationship between the source and target process can be more accurately estimated.

Under the null model which accounts for the periodic structure, the remaining deviation between the observed and expected cross-correlogram corresponds to the induced interaction, which appears as a localized peak at positive lag. This demonstrates that, rather than removing temporal structure from the data, incorporating it into the null model preserves interpretable baseline behavior while isolating deviations indicative of interaction.

\section{Evaluation and Comparison}
\label{sec:validation}

We have outlined several methods for identifying temporal dependencies between event processes while accounting for co-movement arising from shared temporal structure. Selecting an appropriate procedure depends on the target setting, the extent of prior knowledge about the functional form of temporal effects, and whether the goal is to remove or account for such effects in the cross-correlogram. A null model which under-represents structure may over-estimate response relationships while over-accounting for structure may lead to false negatives. In this section we evaluate each method under homogeneous intensity and periodic non-stationarity, demonstrating that detection power depends on whether the null model matches the data generating process. \autoref{app:bursty} assesses method robustness when the generating process is bursty and cannot be reliably parameterized.

We simulate a source and a target process from a known generating intensity for $T=336$ hours. All simulations use cross-correlograms with a half-width $w=3$ hours, bin width $0.5$ hours and each value is calculated across 1000 iterations. Three settings are used. The first is a homogeneous Poisson process with constant intensity function. 
The second is an inhomogeneous Poisson process with a unimodal circadian intensity,
\begin{equation}
\lambda(t) = 5.35 + 5\sin\left(\frac{2\pi (t - \varphi)}{24}\right).
\end{equation}
The third is an inhomogeneous Poisson process with the bimodal circadian intensity, 
\begin{align*}
    \lambda(t) = 5.35\bigg[1 +& 0.45\cos\left(\frac{2\pi(t-\varphi-15)}{24}\right) \\ &+ 0.3\cos\left(\frac{4\pi(t-\varphi-9}{24}\right) \bigg].
\label{eq:bimodal}
\end{align*}
$\varphi$ gives the phase offset between source and target. Each of these three intensity functions shares the mean rate of $5.35$ events per hour. 

Source and target differ only in phase, with a $3$ hour delay in both the unimodal and bimodal settings. The unimodal periodic intensity for the source has a peak at 9:00 and the bimodal has peaks at 10:30 and 19:30, resembling the morning and evening activity peaks reported for online behavior \citep{aledavoodChannelSpecificDailyPatterns2016}.

A dependency between the simulated processes is induced by selecting a proportion $\rho$ of source events and inserting a delayed copy of each into the target, with delays drawn from $\mathcal{N}(1, 0.3^{2})$. Detection probability is measured as a function of $\rho$. At $\rho = 0$ the two processes are independent, so the detection probability there is the empirical rejection frequency.

\begin{figure*}[t]
    \centering
    \includegraphics[width=\textwidth]{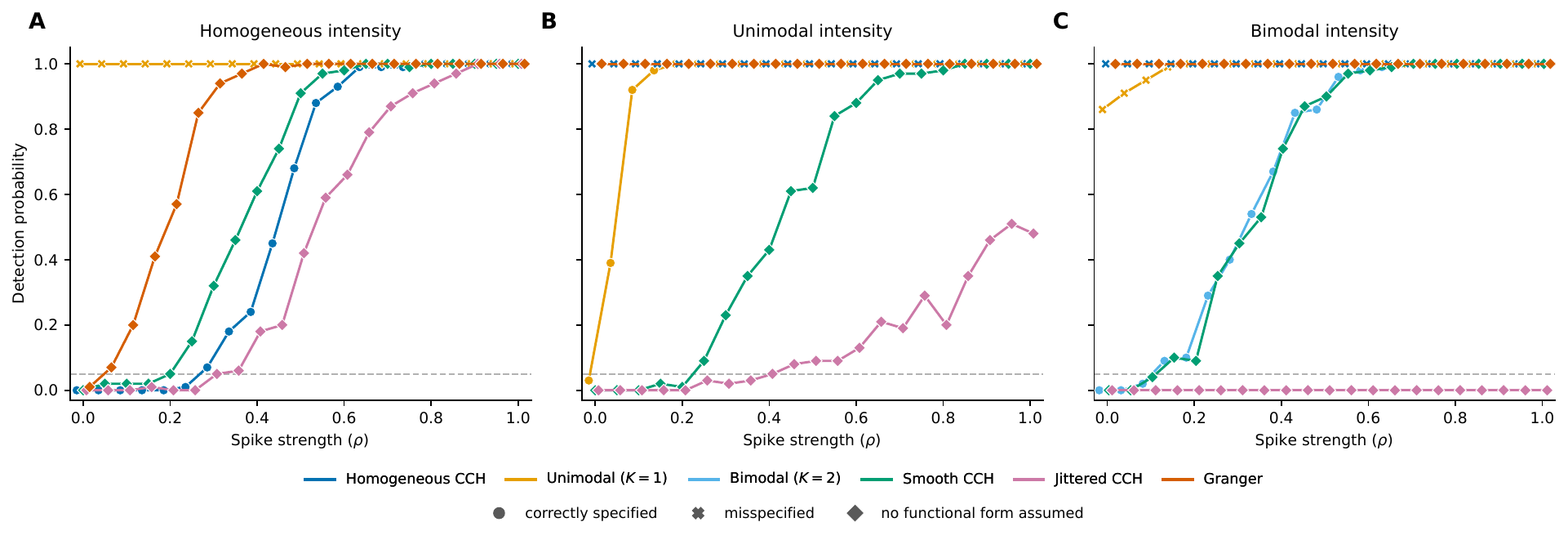}
    \caption{Detection performance as a function of interaction strength for simulated point processes generated under (A) homogeneous and (B) periodic $K=1$ and (C) periodic $K=2$ intensity settings. In each simulation, a proportion $\rho$ of source events is selected and a delayed copy is added to the target process. Delays are drawn from a normal distribution $\mathcal{N}(1,0.3)$, producing a response profile or spike with a Gaussian shape. Source to target interaction dependence increases with $\rho$, with $\rho=0$ representing a system with an independent source and target. Each curve shows the proportion of simulations in which a significant interaction is detected for a given data setting and cross-correlogram method, based on 1000 simulations at each value of $\rho$. Mismatched null models and methods show a detection probability that does not depend on $\rho$. Some remain near 1.0 regardless of interaction strength (homogeneous cross-correlograms and Granger causality for periodic data). In the homogeneous setting (A), Granger causality outperforms cross-correlogram methods, achieving a detection sensitivity close to 1 around $\rho=0.35$. The smooth null model has a higher detection probability than the correctly matched functional homogeneous null across all values of $\rho$, potentially due to its ability to accommodate small rate fluctuations. The jitter procedure does not detect a significant spike until $\rho=0.35$, failing to detect some spikes until $\rho=0.95$. In panel (B), the functional $K=1$ and smooth-intensity nulls account well for the periodic structure, achieving high detection power with low false-positive rates with the correctly specified unimodal null performing best. Panel (C) shows similar performance for smooth and periodic $K=2$ null models. The mismatched $K=1$ model falsely rejects the null hypothesis. Values in \autoref{tab:results}.}
    \label{fig:nonburstyresults}
\end{figure*}

Formally, let $\{t_{x_1}, \ldots, t_{x_{N_X}}\}$ denote the event times of the source, as in \autoref{eq:intensity_form}. Each source event independently generates a delayed response in the target, with $Z_j \sim \mathrm{Bernoulli}(\rho)$ indicating whether event $t_{x_j}$ generates one and $D_j \sim g$ gives the associated delay. The target process is the sum of a baseline process with intensity $\lambda_Y^0(t)$ and the contribution of induced events $\{x_i + D_i \mid Z_i = 1\}$. The expected conditional intensity of the target given the source is
\begin{equation}
\mathbb{E}\left[\lambda_Y(t) \mid X\right]
= \lambda_Y^{0}(t) + \rho \sum_{j=1}^{N_X} g\!\left(t - t_{x_j}\right),
\label{eq:artificial_spike}
\end{equation}
where $g$ is the density of the delay distribution. In this section, $g$ is set to $\mathcal{N}(1, 0.3^2)$. When $g$ integrates to one, the expected number of induced events is $\rho N_X$, so $\rho$ directly controls the strength of the induced interaction between the processes.

To evaluate how detection probability is affected by the fit between the functional form of the smooth null model and the true generating process, we use a family of periodic null models which can be changed from the unimodal to bimodal case. 

The \textit{homogeneous null} assumes a constant rate. The \textit{periodic null} is a trigonometric polynomial of order $K$,
\begin{equation}
\lambda_K(t) = c_0 + \sum_{k=1}^{K}
  \left[\,c_k \cos\!\left(\frac{2\pi k t}{P}\right)
        + s_k \sin\!\left(\frac{2\pi k t}{P}\right)\right].
\label{eq:harmonic}
\end{equation}
At $K = 1$ this is the unimodal model, one standard parametric description of a circadian effect sometimes called the cosinor \cite{cornelissen2014cosinor}. At $K=2$ we have a bimodal periodic model. 

To test the relationship between performance and fit of the functional model, we will test how well the periodic null with $K=1$ and $K=2$ performs in a bimodal periodic data setting. As the periodic null has an additive structure, we know the variance of the periodic fluctuation due to the second term ($k=2$) is $30.77\%$ for the parameters used in these simulations.

The \textit{hour-of-day null} averages the rate for each period (here $2$ hours) across all period in the dataset, using multiple days of data to improve fit when periodic structures are consistent across the process. In the homogeneous case without periodic structure, a moving average baseline replaces the hour-of-day null. We also use interval jitter and Granger causality computed on binned counts as baseline measures. 

In this section we run two main experiments. We compare the performance of the exact functional form of the generating process against other null models and baseline methods and assess how the functional null model performs when the structure is misspecified. For the unimodal and bimodal settings we use the exact generating intensity where possible, estimating parameters for mismatched or non-parametric models.
Across both experiments, to evaluate whether a spike was detected, we employ a significance test which uses a scan statistic \cite{kulldorff1997spatial}, defined as the maximum contiguous accumulation of positive deviations in the whitened cross-correlogram. This test evaluates whether a contiguous set of bins exhibits a jointly elevated count relative to the null expectation, with statistical significance assessed by Monte Carlo calibration under the null model. The procedure is similar to cluster-based non-parametric testing applied in neurophysiology by Maris and Oostenveld \cite{maris2007nonparametric}. 
\\
This significance testing procedure is repeated for each set of simulated event-times across each method for $1000$ iterations, with the reported detection probability, given by the proportion of times a significant spike was detected for each value of $\rho$ using a significance level of $\alpha = 0.05$.  

These results are summarized in \autoref{fig:nonburstyresults} for homogeneous and periodic settings, with complete results in \autoref{tab:results}. \autoref{app:bursty} evaluates robustness under bursty dynamics, where the temporal structure cannot be specified functionally.
\\
Hyperparameters for the smooth intensity function and jittering windows were selected on a coarse grid (0.25, 0.5, 1, 3, 6, 12, 24 hours). Initial selection used $\rho \in \{0, 0.1, 0.2, 0.3\}$ at 50 iterations, then the grid was reduced and run for each synthetic data setting across 1000 iterations. Hyperparameter selection occurs first by comparing the distance between the fitted null intensity and the known generating intensity function to reduce the potential parameter set. For remaining values we compute empirical size and power. Among settings whose empirical size at $\rho = 0$ is compatible with $\alpha$ after allowing for Monte Carlo error, we choose the setting with the greatest power at low values of $\rho$. 
Shimazaki and Shinomoto \citep{shimazaki2010kernel} describe principled selection criteria for kernel-based rate estimation. Final hyperparameter values were a $6$ hour window for the interval averaging in the homogeneous setting, a $2$ hour window for the unimodal and bimodal smooth null, jitter windows of $0.5$ and $12$ hours for the homogeneous and unimodal settings and a bin size of $1$ hour for Granger causality.

As interaction strength increases (increasing $\rho$), detection probability rises when the null model matches the temporal structure of the data generating process, i.e. the homogeneous cross-correlogram for homogeneous data and the correctly specified periodic cross-correlograms for periodic data. When the periodic structure is not specified correctly, i.e. $K=1$ for the bimodal setting (\autoref{fig:nonburstyresults}C), the method incorrectly detects a response at $\rho=0$ 86\% of the time and is invalid.

For the periodic settings, the smooth null model exploits the periodic structure by averaging the rate for each hour across all days, producing a more stable rate estimate. This method performed well across both unimodal and bimodal settings, outperforming the correctly specified null in the bimodal setting. In the unimodal setting it was less sensitive than the correctly specified periodic $K=1$ null, and did not detect a significant spike until $\rho \geq 0.3$. 

In contrast, mismatched null models produce spurious detections or reduced sensitivity. Homogeneous null models applied to periodic data consistently detected significant spikes even when artificial spikes were small or absent (low $\rho$). For the homogeneous setting, the 24-hour jitter interval had lower performance than the other cross-correlogram based methods, but all were outperformed by Granger causality. In both periodic settings, the true positive rate of Granger causality does not correlate with $\rho$.

In the appendices \autoref{fig:compperiodic} and \autoref{fig:compjitter} show the resulting cross-correlogram (blue) and theoretical null model (magenta) for a single simulation of an inhomogeneous Poisson process with a periodic or bursty intensity function respectively. Panel (ii) shows the functional model fit used for the periodic data setting for these figures. Together \autoref{fig:compjitter} and \autoref{fig:compperiodic} demonstrate how when the null model matches the data-generating process, detection power increases reliably with interaction strength, whereas mismatched nulls lead to either false positives or reduced sensitivity.

\subsection{\bf Practical Procedure for Implementation}

The preceding analysis examined the performance and limitations of the cross-correlogram under controlled conditions. We now extend this framework to real-world data, where multiple interacting processes may contribute to observed temporal dependencies. In such settings, a systematic approach is needed to identify and interpret significant pairwise relationships. Because the cross-correlogram characterizes dependencies as a function of lag, multiple features of its structure must be evaluated to detect significant temporal interactions. We propose a systematic framework for applying cross-correlogram methods to detect and quantify these relationships.

The following steps outline a recommended workflow for applying cross-correlogram methods to empirical data:
\begin{enumerate}
\item \textbf{Data preparation and suitability assessment.}
Ensure the dataset contains sufficient events to support statistical inference. Cross-correlogram significance tests generally require a minimum of approximately five counts per bin; the required number of events depends on event frequency and the response window of interest.

\item \textbf{Correction for unwanted temporal effects.}
Select and apply an appropriate method to account for or remove systematic temporal structure unrelated to the target relationship. The theoretical and experimental comparisons in \autoref{sec:validation} may guide this choice.

\item \textbf{Cross-correlogram construction.}
Compute the cross-correlogram between event time series, ensuring consistent binning and normalization across analyses.

\item \textbf{Significance testing.}
Apply statistical tests to evaluate either (1) the overall structure of the cross-correlogram or (2) the significance of individual bins. A two-sided per-bin Poisson test or equivalent may be used, with multiple-comparison correction where appropriate.

\item \textbf{Quantification of temporal relationships.}
Extract interpretable summary measures such as per-bin significance values, overall correlation strength, error metrics, or the temporal offset corresponding to the weighted mean of the cross-correlogram.
\end{enumerate}

Temporal structure can be removed from the data, i.e. interval jitter, or represented in the null model, i.e. smooth intensity null. The choice depends on whether the timescale of the unwanted structure is separated from the timescale of the dependency of interest. Where the separation is clear, either approach recovers the dependency and interval jitter is the cheaper option. Where the scales overlap, only a smooth null can preserve target dependencies.

When the goal is to isolate specific delayed interactions, removing temporal structure is advantageous. Conversely, when retaining the full temporal structure is desirable but significance must be assessed against a modeled baseline, a modified null model (e.g., smooth intensity or periodic function) may be preferable.

If the functional form of unwanted temporal structure can be parameterized, a smooth parametric model can account for these effects directly within the cross-correlogram. This strategy is particularly effective when parameters can be estimated from aggregated data subsets, reducing the risk of overfitting in small-sample settings.

When the characteristic timescales of unwanted and target effects differ substantially, non-parametric smooth intensity functions or interval-jitter methods can effectively distinguish between them. The choice depends on whether the analysis aims to remove or explicitly represent these temporal patterns.

\section{Application: Temporal Interaction Networks in Social Media}
\label{sec:appdata}

Social data commonly exhibit strong temporal structure and intermittent burstiness. Human behavior on social platforms shows circadian, weekly and seasonal patterns, as well as transient surges of collective attention \citep{kobayashiModelingCollectiveAnticipation2021,barabasi2005origin,sanliLocalVariationHashtag2015}. Crucially for temporal inference, many social interactions occur at time scales (minutes to weeks) that overlap with these periodic and seasonal effects, for example collective social action like the Black Lives Matter protest, which became viral across social media over the period of several days \citep{bestvaterTenYearsBlackLivesMatter2023}. As shown in \autoref{sec:validation} and \autoref{app:bursty}, such non-stationary and bursty dynamics can induce structured deviations in cross-correlograms that mimic pairwise interactions; below we apply the cross-correlogram with a smooth inhomogeneous null to a large social media dataset and demonstrate how the method highlights delayed temporal relationships that are not captured by simple co-occurrence.

Co-occurrence is a commonly used measure to detect coordinated activity on social networks, motivated by the hypothesis that interactions occurring extremely quickly are extremely unlikely to have occurred organically, and so, similar activity occurring within short-timescales is commonly used to detect coordinated inauthentic activity \cite{grahamVirusCoordinatedSpread2020}. This method also has the advantage of being simple to understand and implement, but is not able to capture the more complex temporal response profiles which the cross-correlogram method aims to capture.

\begin{figure*}[t]
    \centering
    \includegraphics[width=\textwidth]{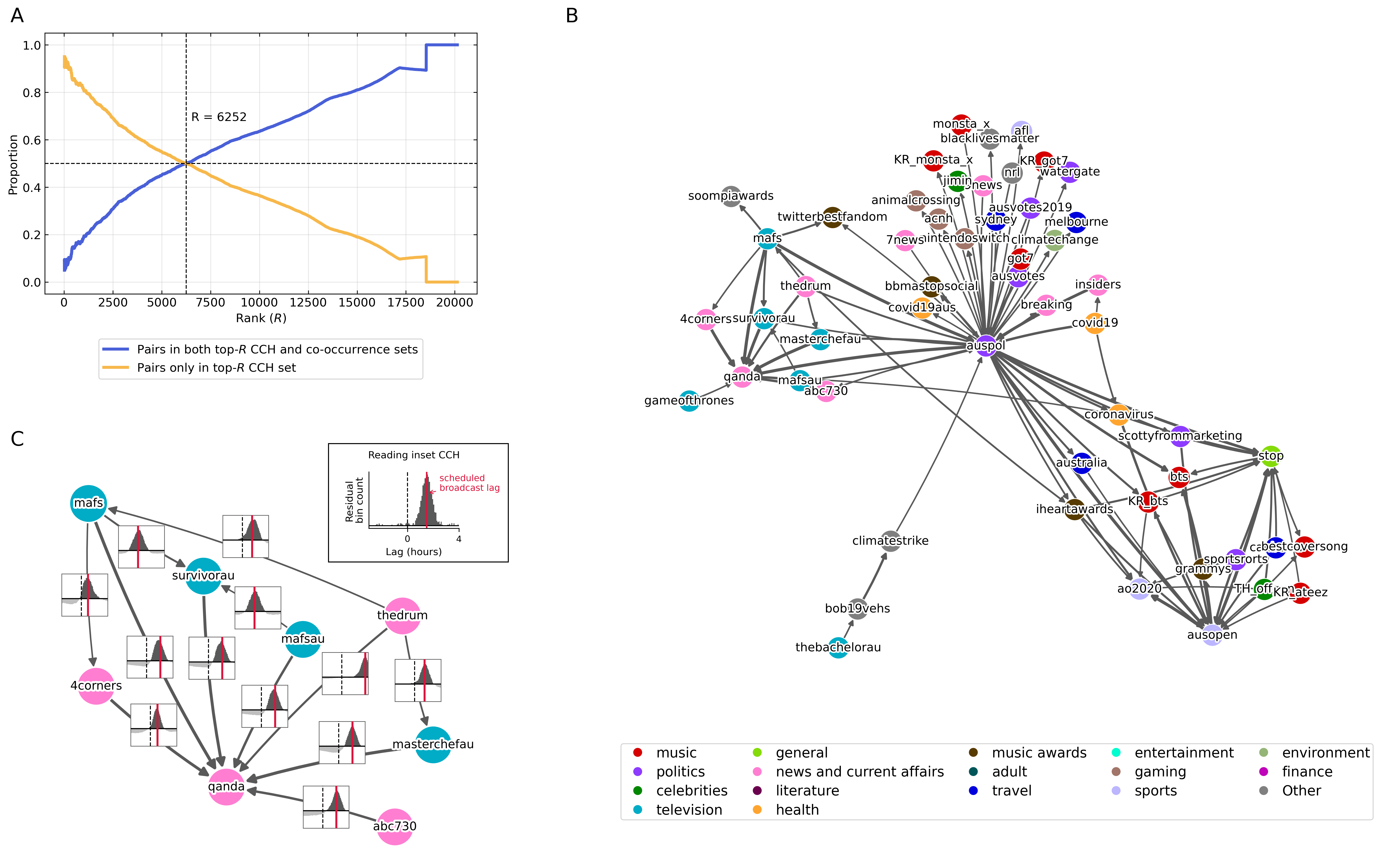}
    \caption{The cross-correlogram (CCH) method identifies temporal relationships that differ from those captured by exact-time co-occurrence. For each rank threshold $R$, we compare the top-$R$ hashtag pairs ranked by CCH strength with those ranked by normalized exact timestamp co-occurrence. CCH strength $s$ is the mean absolute deviation of the observed cross-correlogram from the smooth-intensity null expectation.
    (A) Agreement between the two rankings as $R$ increases. The blue curve shows the proportion of top-$R$ CCH edges which also appear among the top-$R$ co-occurrence edges. The orange curve shows the proportion unique to the top-$R$ CCH ranking. Low agreement for small to moderate $R$ indicates that highly ranked delayed temporal relationships often differ from frequently co-occurring pairs. 
    (B) Network formed from the top ranked CCH edges that are absent from the top-ranked co-occurrence network ($R=100)$. Nodes represent hashtags colored by topic. Hashtags originally written in non-Latin scripts are transliterated into English and prefixed by a two-letter language code (e.g., 에이티즈 becomes KR\_ateez). Directed edges indicate predominant temporal ordering with width proportional to $s$. A force-directed layout reveals structured groups of delayed relationships around topics including Australian politics and news, television, sport and music fandom, with \#auspol acting as a hub.
    (C) Television and current-affairs subgraph from (B), with residual cross-correlograms shown on each edge. Dashed black lines indicate zero lag and red lines the difference in scheduled broadcast times during 2019–2020. CCHs use an extended (-3,+4) hour window for visualisation and 5-minute bins. Peaks near scheduled lags are consistent with delayed relationships associated with broadcast order.}
    \label{fig:four_subfigures_swapped}
\end{figure*}

\subsection{Dataset and Preprocessing}
We analyze a collection of posts from \textsc{X} (formerly Twitter) covering January 2019 to September 2020 available at \cite{AUDATAMitchell2026}. The raw dataset contains approximately 17.1M tweets matching a set of Australian location keywords\footnote{Keywords taken from common state and territory names (including abbreviations) and the names of the largest Australian cities, towns and regions by population. The full list is: `australia',`adelaide', `perth',`melbourne', `darwin', `launceston', `hobart', `canberra', `sydney', `brisbane', `cairns', `wollongong', `geelong', `townsville', `toowoomba', `ballarat', `victoria', `queensland', `tasmania', `bendigo',`wodonga', `gold coast', `sunshine coast', `tweed heads', `queanbeyan', `maitland', `nsw', `qld'}. After extracting hashtags and retaining the top 200 most frequent hashtags, the processed dataset contains $3{,}133{,}247$ hashtag-time pairs (median events per hashtag: 10,392). Hashtags were assigned topics and subtopics using ChatGPT-4o \citep{openai_chatgpt_2023} followed by manual review, producing 26 unique topic labels. Topics containing at least 5 unique hashtags were retained and an `Other' topic was used for any topics which contained fewer than 5 unique hashtags. Occasional noise (e.g.\ ``Victoria'' in non-Australian contexts) remains but does not affect the qualitative results; the full mapping and prompt used are provided in \autoref{app:topicclassifications}.

We represent each hashtag as a point process of event times and compute pairwise cross-correlograms for each ordered pair using the smooth null model described in \autoref{sec:SIF}. For the smooth null we used an interval-averaged intensity estimator with an averaging window of 6 hours; the cross-correlogram window was set to $\pm 3$ hours with bin width $\Delta = 5$ minutes. These choices ensure the null captures variability on scales $\geq 6\,$h while preserving sensitivity to interactions on shorter scales (minutes–hours), the temporal range typically of interest in social coordination. 

Alongside the cross-correlogram network, we construct an exact-time co-occurrence network, where the edge weight between hashtags i and j, $w_{ij}^{\text{co}}$, is the proportion of their combined unique timestamps at which both hashtags occur,
$$
w_{ij}^{\text{co}} = \dfrac{N_{ij}}{N_i+N_j-N_{ij}}
$$
where $N_{ij}$ is the number of exact-timestamp co-occurrences and $N_i$ and $N_j$ are the respective timestamp frequencies of the two hashtags. 

Conventional analyses of social media coordination rely on co-occurrence within fixed windows \citep{grahamVirusCoordinatedSpread2020}, which capture immediate synchrony but overlook delayed responses. The cross-correlogram framework is complementary and can identify pairwise relationships that are driven by delayed, structured temporal responses rather than simultaneous event times.

To analyse these networks, we use a network comparison technique similar to the rank-based overlap proposed by Webber et al. \cite{webber2010similarity}. For each of the cross-correlogram and co-occurrence networks, edges are ranked by their edge weights, with ties assigned the average rank. To identify relationships preferentially detected by the cross-correlogram measure, we compare the top-$R$ hashtag pairs ranked by CCH strength with the top-$R$ pairs ranked by exact-timestamp co-occurrence at each rank threshold $R$. This comparison quantifies agreement between the two measures and identifies relationships detected by the cross-correlogram procedure but not by exact-timestamp co-occurrence. We subsequently construct a differenced network from these relationships.

All analyses use a cross-correlogram window of $\pm3$ hours. For visualisation only, the residual cross-correlograms displayed in \autoref{fig:four_subfigures_swapped} (C) are shown over $(-3,+4)$ hours to include the full peak of more delayed responses; this extended display window does not affect the underlying analysis.

\subsection{Results and Network Analysis}
\autoref{fig:four_subfigures_swapped}A shows agreement between the cross-correlogram and exact-timestamp co-occurrence rankings across each rank threshold $R$, together with relationships only present in the top-$R$ cross-correlogram relationships. The resulting differenced network for $R=100$ is shown in \autoref{fig:four_subfigures_swapped}B. Each node is a hashtag colored by topic and node size reflects overall frequency. \autoref{fig:four_subfigures_swapped}C shows a subset of this network with the corresponding residual cross-correlogram overlaid above each edge. The selected cluster contains nodes which represent news, current affairs and entertainment television programs. This subset provides an interpretable case in which known broadcast delays can be compared with the temporal response delays estimated by the cross-correlogram procedure.

Let $r_j = h_j - \hat h_j$ denote the residual in bin $j$. For each ordered pair we compute
\begin{equation}\label{eq:s_d}
    s = \frac{1}{N_b}\sum_{j=1}^{N_b} \left| r_j \right|,
\qquad
    d = \frac{\sum_{j=1}^{N_b} b_j\, r_j}{w \sum_{j=1}^{N_b} \left| r_j \right|},
\end{equation}
where $h_{j}$ is the observed count in bin $j$ which is given by the number of intra-process event times in $[b_j,b_j+\Delta)$. $\hat{h}_{j}$ is the theoretical (expected) count under the smooth inhomogeneous Poisson null, $N_b$ is the total number of bins in the cross-correlogram and $w$ is half of the overall cross-correlogram window, so that $d\in[-1,1]$. Intuitively, $s$ measures the average absolute deviation between observed and expected bin count for bin $j$ (a measure of disagreement between data and null), while $d$ summarizes the direction of deviations (positive $d$ indicates a tendency for $X$ to precede $Y$). In the cross-correlogram network, edge weights are given by $s$ and their direction by the sign of $d$.

In \autoref{fig:four_subfigures_swapped}B edges are represented with weight proportional to $s$ (darker = larger deviation from the expected cross-correlogram) and node placement is force-directed with attraction according to edge weight.
The agreement between the top pairwise relationships detected by the cross-correlogram and co-occurrence methods is low for even moderate values of $R$, reaching 50\% agreement only once the top $R=6,252$ pairs are included by each measure (31\% of all 19,900 unordered hashtag pairs).

\autoref{fig:four_subfigures_swapped}B shows the differenced top-100 edge network. This network contains pairwise relationships with the top-$100$ residuals under the cross-correlogram method which do not appear among the top-$100$ exact-timestamp co-occurrence pairs. These are the pairwise relationships which a co-occurrence analysis would be less likely to detect and the resulting network shows thematic clusters around topics such as Australian politics and news, television, sport, and music fandom connected through \#auspol, which acts as a hub. The relationships visible in the resulting network reflect ongoing events, such as voting campaigns for music awards or hashtags associated with particular sporting or political events.

\autoref{fig:four_subfigures_swapped}C focuses on hashtags linked to Australian television programs (\#mafs, \#mafsau, \#survivorau, \#masterchefau) and news and current affairs television programs (\#4corners, \#qanda, \#abc730, \#thedrum). The detected response between each pair of programs from the cross-correlogram method matches with the known broadcast delay shown with a red line on the residual cross-correlograms\footnote{Broadcast times (regular schedule) are Married at First Sight Australia (\#mafs, \#mafsau) at 19:30; Survivor Australia (\#survivorau) at 19:30; Masterchef Australia (\#masterchefau) at 19:30; Four Corners (\#4corners) at 20:30; Q+A (\#qanda) at 21:35; The Drum (\#thedrum) at 18:00; and 7.30 (\#abc730) at 19:30.}. These examples demonstrate the ability of the smooth-null cross-correlogram to reveal delayed, event-driven relationships within broader  background synchrony.

Network analysis highlights the differences between the cross-correlogram and co-occurrence networks. At the level of $R=100$ shown in \autoref{fig:four_subfigures_swapped}B, the co-occurrence network is sparse with its largest connected component containing only 9 of 96 active nodes, compared to 55 of 57 in the cross-correlogram network. At $R=100$ the networks are sparse, to compare connectivity and clustering we set $R=500$. At this level \#auspol is the most connected node in the cross-correlogram network (degree 136, local clustering co-efficient 0.034) with a degree of 22 in the co-occurrence network. In the data \#auspol is the most commonly occurring hashtag in the dataset with $343,972$ occurrences, with the second most commonly occurring \#bts occurring $99,148$ times. This may lead to an outsized degree in the cross-correlogram network which is not corrected by occurrence rate.

At $R=500$ the cross-correlogram network is fully connected with a global clustering coefficient of 0.109 and Louvain community detection finds four non-trivial communities. The largest connected component of the co-occurrence network contains 153 of 181 active hashtags with a global clustering coefficient of 0.613. Using Louvain community detection, the co-occurrence network contains 27 non-trivial communities, ranging from 2 to 30 hashtags, while the cross-correlogram network contains 4 communities ranging in size from 4 to 111 nodes. Nodes of highest eigenvector-centrality in the cross-correlogram network are \#stop, \#ausopen, and \#ao2020. In the co-occurrence network, the highest-centrality hashtags are \#asmsg, \#spub, \#iartg, \#kindle, and \#ibooks, a group associated with self-publishing and book promotion.

The application demonstrates that the cross-correlogram with smooth null model can capture meaningful delayed temporal relationship that are missed by simple co-occurrence measures. As a threshold-free complement to \autoref{fig:four_subfigures_swapped}A, the Spearman rank correlation between CCH strength $s$ and exact-timestamp co-occurrence across all 19,900 hashtag pairs is weak (Spearman rank correlation of $0.379$, $p < 10^{-300}$), confirming that the two measures order pairwise relationships quite differently across the full dataset and not only among their most strongly ranked edges. Together, these results illustrate how the cross-correlogram framework provides interpretable, data-driven temporal insights into collective online behavior.
\FloatBarrier
\section{Discussion and Conclusion}
\label{sec:discussion}

Detecting directional temporal relationships in social data is challenging due to the noisy, bursty, and non-stationary nature of these systems. Common tools, such as co-occurrence analysis, Granger causality, and transfer entropy, capture aspects of temporal dependence but their applicability is limited when applied to irregular, discrete event streams. In this work, we introduced the cross-correlogram framework as an interpretable, event-based approach for estimating lagged dependencies directly from such data and demonstrate its utility on both simulated and real data.

Our framework extends classical cross-correlogram analysis by incorporating smooth, inhomogeneous null models that account for known behavioral rhythms and time-varying event rates typical of social data. We analytically derived the bias introduced by bursty and periodic effects under homogeneous Poisson assumptions, derived the additional error introduced by correcting for them with interval jitter, proposed functional and empirical intensity models that instead represent them in the null, and validated these through simulation. Under known periodic structure the functional model recovers the dependency profile. When the structure cannot be parameterized, the empirical model retains sensitivity to true interactions (\autoref{app:bursty}). The smooth-null formulation enables realistic significance testing without perturbing event times, bridging data-driven inference with behavioral modeling.

Applied to 3.1 million social-media event times, we demonstrate how a smooth-null cross-correlogram reveals interpretable temporal relationships which complement those detected by co-occurrence methods. The smooth-null cross-correlogram uncovered delayed, asymmetric relationships between hashtags, distinguishing brief event-driven bursts from sustained synchrony. The cross-correlogram analysis with the smooth intensity function identified delays between television programs which matched published broadcast times, providing external validation that the recovered lags reflect real, structured behavior rather than artifacts of shared attention cycles.

Despite its strengths, several limitations remain. First, the current formulation assumes Poisson event generation, which may under represent higher-order dependencies or self-exciting dynamics characteristic of social activity. Extending the smooth-null approach to Hawkes or renewal processes could capture such feedback explicitly. Second, computing all pairwise cross-correlograms scales quadratically with the number of processes, motivating approximate or sparse representations for large-scale systems. Third, while the method infers directional timing relationships, it does not establish causality; this distinction requires explicit mechanistic modeling.

Future work will pursue hierarchical and multivariate extensions to capture community-level temporal dependencies, and adaptive smoothing schemes that learn intensity variation directly from data rather than fixed functional forms. Beyond methodological development, the interpretability of lag profiles and their visualization as temporal networks offer promising avenues for studying collective attention, information diffusion, and coordination in social and other complex systems.

By combining classical temporal-correlation analysis with modern functional modeling, the cross-correlogram framework provides a unified approach for quantifying delayed dependencies in bursty, non-stationary data. Its theoretical grounding, validated performance, and empirical versatility suggest broad potential across domains where repeated trials are rare but rich event-time data abound. More broadly, this work demonstrates how embedding known behavioral rhythms within statistical null models can substantially improve the robustness and interpretability of temporal inference in complex systems.

\section{Acknowledgments}
The authors gratefully acknowledge Lewis Mitchell for providing the data used in this study.

\section{Funding}
This work is supported in part by funds from the National Science Foundation (NSF: \# 1636933 and \# 1920920).
RL acknowledges support from the EPSRC Grants EP/V013068/1, EP/V03474X/1, and EP/Y028872/1.
T.A. is supported by JSPS KAKENHI Grant Number JP24K07699, and JSPS KAKENHI Grant Number JP24K03007. 
R.K. is supported by JSPS KAKENHI  JP21H03559, JP22H03695, and JP23K24950, JST FOREST JPMJFR232O, and AMED JP223fa627001.

\section{Author contributions statement}
B.S., T.A., and R.K. conceived the project. B.S. wrote the analysis code, performed the numerical simulations and data analyses, and developed and implemented the methodology. 
R.L., T.A., and R.K. supervised the project. B.S. drafted the manuscript with input from all authors.

All authors provided input on the methodology, read and approved the final manuscript.

\section{Data availability}
The X / Twitter dataset was provided by \cite{AUDATAMitchell2026} and is available at \url{https://adelaide.figshare.com/articles/dataset/Australian_Twitter_Dataset_2019-20/33283899}.

\bibliographystyle{plain}
\bibliography{HashtagDynamics}

\FloatBarrier
\begin{appendices}

\newpage
\section{Comparison of methods on periodic data}

\begin{figure}[t]
    \centering
    \includegraphics[width=0.5\textwidth]{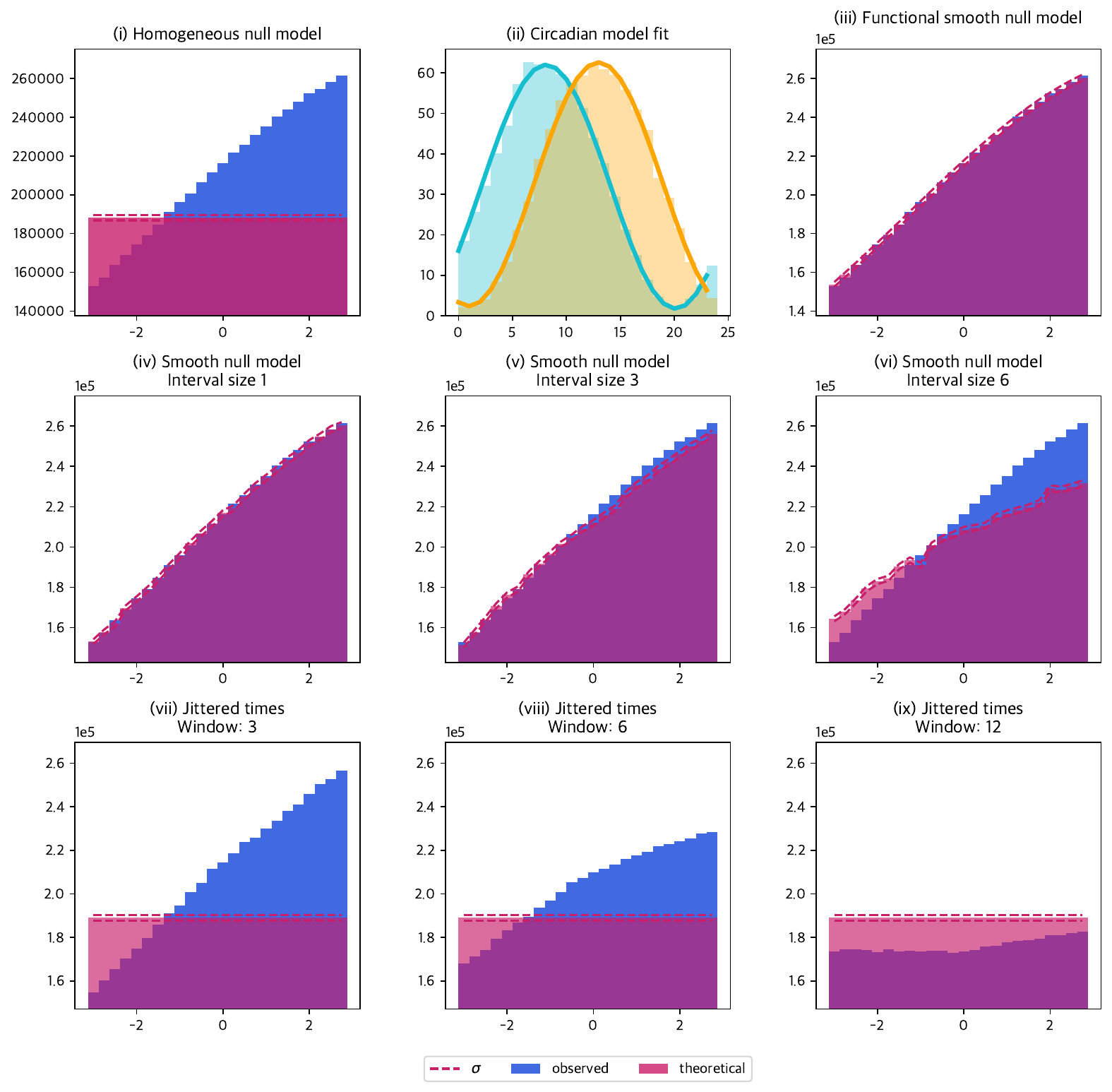}
        \caption{\small Various cross-correlograms constructed using simulated data with a periodic intensity function. The offset in intensity function between the source and target process is 4 units. In (i), a constant Poisson null model is used to construct theoretical bin heights shown in red. In (ii), the times over a cycle length of 24 are used to fit a periodic model to the data. This model is used to construct a smooth null model with a functional form, shown in (iii). This functional model is able to account for the form of the cross-correlogram, revealing the spike can be accounted for using the specified periodic model. Subplots (iv) - (vi) show a smooth null model which uses an interval average to account for changes in the underlying intensity function of the target and the source process. The interval size needs to be chosen to mitigate unwanted temporal effects. In this simulated data, the temporal effect has length 24, and this approach begins to fail between an interval smoothing window of 3 and 6. However, (iv) shows that an interval size of 1 and 3 can account for fluctuations on a 24 hour scale well, with some deviations from the observed bin heights. In (vii)-(ix) show the behaviour when interval jitter is applied to the source and target times. As the interval size increases, the temporal structure is removed from the observed cross-correlogram, and with a window of 24 this structure is almost completely removed.}
    \label{fig:compperiodic}
\end{figure}

\autoref{fig:compperiodic} shows the resulting cross-correlogram (blue) and theoretical null model (magenta) for a single simulation of an inhomogeneous Poisson process with a periodic intensity function.

Each panel shows a different null model, showcasing the key differences between the methods. Panel (ii) shows the functional model fit used for the periodic data setting across all simulations.

\autoref{fig:compperiodic} demonstrate how when the null model matches the data-generating process, detection power increases reliably with interaction strength, whereas mismatched nulls lead to either false positives or reduced sensitivity.

In jitter-based cross-correlograms, visual inspection alone often fails to determine statistical significance and to distinguish null-model artifacts from structure at specific temporal scales.When sufficient data are available, smooth functional null models provide clearer identification of meaningful deviations from theoretical expectations.

\begin{table*}
\caption{Detection probabilities (proportion of simulations with $p<0.05$) as a function of induced spike strength $\rho$. Results are shown by data generating process (column groups) and by null model or method (subcolumns). The smoothing or jittering interval was selected for each data generating process and the corresponding hyperparameter is listed in each relevant column. Note H. represents the homogeneous null model with constant intensity.}
\label{tab:results}
\scriptsize
\tabcolsep=2pt
\begin{tabular*}{\textwidth}{@{\extracolsep{\fill}} p{1cm} |
    c c c c c |
    c c c c c |
    c c c c c c
  @{}}
\hline
 & \multicolumn{5}{c}{\textbf{Homogeneous data}}
 & \multicolumn{5}{c}{\textbf{Periodic data ($K=1$)}}
 & \multicolumn{6}{c}{\textbf{Periodic data ($K=2$)}} \\
\cmidrule(lr){2-6} \cmidrule(lr){7-11} \cmidrule(lr){12-17}
$\;\rho\;$
 & H. & Unimodal & Smooth & Jittered & GC
 & H. & Unimodal& Smooth & Jittered & Granger
 & H. & Unimodal & Bimodal & Smooth & Jittered & Granger
\\
 & & $K=1$ & ($6$ hour) & ($0.5$ hour) &
 & & $K=1$ & ($2$ hour) & ($12$ hour) &
 & & $K=1$ &$K=2$  & ($2$ hour) & ($24$ hour) & \\
\hline

0.0 & 0.0 & 1.0 & 0.0 & 0.0 & 0.01 & 1.0 & 0.03 & 0.0 & 0.0 & 1.0 & 1.0 & 0.86 & 0.0 & 0.0 & 0.0 & 1.0 \\
0.05 & 0.0 & 1.0 & 0.02 & 0.0 & 0.07 & 1.0 & 0.39 & 0.0 & 0.0 & 1.0 & 1.0 & 0.91 & 0.0 & 0.0 & 0.0 & 1.0 \\
0.1 & 0.0 & 1.0 & 0.02 & 0.0 & 0.2 & 1.0 & 0.92 & 0.0 & 0.0 & 1.0 & 1.0 & 0.95 & 0.02 & 0.04 & 0.0 & 1.0 \\
0.15 & 0.0 & 1.0 & 0.02 & 0.01 & 0.41 & 1.0 & 0.98 & 0.02 & 0.0 & 1.0 & 1.0 & 0.99 & 0.09 & 0.1 & 0.0 & 1.0 \\
0.2 & 0.0 & 1.0 & 0.05 & 0.0 & 0.57 & 1.0 & 1.0 & 0.01 & 0.0 & 1.0 & 1.0 & 1.0 & 0.1 & 0.09 & 0.0 & 1.0 \\
0.25 & 0.01 & 1.0 & 0.15 & 0.0 & 0.85 & 1.0 & 1.0 & 0.09 & 0.03 & 1.0 & 1.0 & 1.0 & 0.29 & 0.35 & 0.0 & 1.0 \\
0.3 & 0.07 & 1.0 & 0.32 & 0.05 & 0.94 & 1.0 & 1.0 & 0.23 & 0.02 & 1.0 & 1.0 & 1.0 & 0.4 & 0.45 & 0.0 & 1.0 \\
0.35 & 0.18 & 1.0 & 0.46 & 0.06 & 0.97 & 1.0 & 1.0 & 0.35 & 0.03 & 1.0 & 1.0 & 1.0 & 0.54 & 0.53 & 0.0 & 1.0 \\
0.4 & 0.24 & 1.0 & 0.61 & 0.18 & 1.0 & 1.0 & 1.0 & 0.43 & 0.05 & 1.0 & 1.0 & 1.0 & 0.67 & 0.74 & 0.0 & 1.0 \\
0.45 & 0.45 & 1.0 & 0.74 & 0.2 & 0.99 & 1.0 & 1.0 & 0.61 & 0.08 & 1.0 & 1.0 & 1.0 & 0.85 & 0.87 & 0.0 & 1.0 \\
0.5 & 0.68 & 1.0 & 0.91 & 0.42 & 1.0 & 1.0 & 1.0 & 0.62 & 0.09 & 1.0 & 1.0 & 1.0 & 0.86 & 0.9 & 0.0 & 1.0 \\
0.55 & 0.88 & 1.0 & 0.97 & 0.59 & 1.0 & 1.0 & 1.0 & 0.84 & 0.09 & 1.0 & 1.0 & 1.0 & 0.96 & 0.97 & 0.0 & 1.0 \\
0.6 & 0.93 & 1.0 & 0.98 & 0.66 & 1.0 & 1.0 & 1.0 & 0.88 & 0.13 & 1.0 & 1.0 & 1.0 & 0.98 & 0.98 & 0.0 & 1.0 \\
0.65 & 0.99 & 1.0 & 1.0 & 0.79 & 1.0 & 1.0 & 1.0 & 0.95 & 0.21 & 1.0 & 1.0 & 1.0 & 0.99 & 0.99 & 0.0 & 1.0 \\
0.7 & 0.99 & 1.0 & 1.0 & 0.87 & 1.0 & 1.0 & 1.0 & 0.97 & 0.19 & 1.0 & 1.0 & 1.0 & 1.0 & 1.0 & 0.0 & 1.0 \\
0.75 & 0.99 & 1.0 & 0.99 & 0.91 & 1.0 & 1.0 & 1.0 & 0.97 & 0.29 & 1.0 & 1.0 & 1.0 & 1.0 & 1.0 & 0.0 & 1.0 \\
0.8 & 1.0 & 1.0 & 1.0 & 0.94 & 1.0 & 1.0 & 1.0 & 0.98 & 0.2 & 1.0 & 1.0 & 1.0 & 1.0 & 1.0 & 0.0 & 1.0 \\
0.85 & 1.0 & 1.0 & 1.0 & 0.97 & 1.0 & 1.0 & 1.0 & 1.0 & 0.35 & 1.0 & 1.0 & 1.0 & 1.0 & 1.0 & 0.0 & 1.0 \\
0.9 & 1.0 & 1.0 & 1.0 & 1.0 & 1.0 & 1.0 & 1.0 & 1.0 & 0.46 & 1.0 & 1.0 & 1.0 & 1.0 & 1.0 & 0.0 & 1.0 \\
0.95 & 1.0 & 1.0 & 1.0 & 1.0 & 1.0 & 1.0 & 1.0 & 1.0 & 0.51 & 1.0 & 1.0 & 1.0 & 1.0 & 1.0 & 0.0 & 1.0 \\
1.0 & 1.0 & 1.0 & 1.0 & 1.0 & 1.0 & 1.0 & 1.0 & 1.0 & 0.48 & 1.0 & 1.0 & 1.0 & 1.0 & 1.0 & 0.0 & 1.0 \\
\hline
\end{tabular*}
\end{table*}

\FloatBarrier

\section{Bursty dynamics}\label{app:bursty}

\begin{figure}[t]
    \centering
    \includegraphics[width=0.5\textwidth]{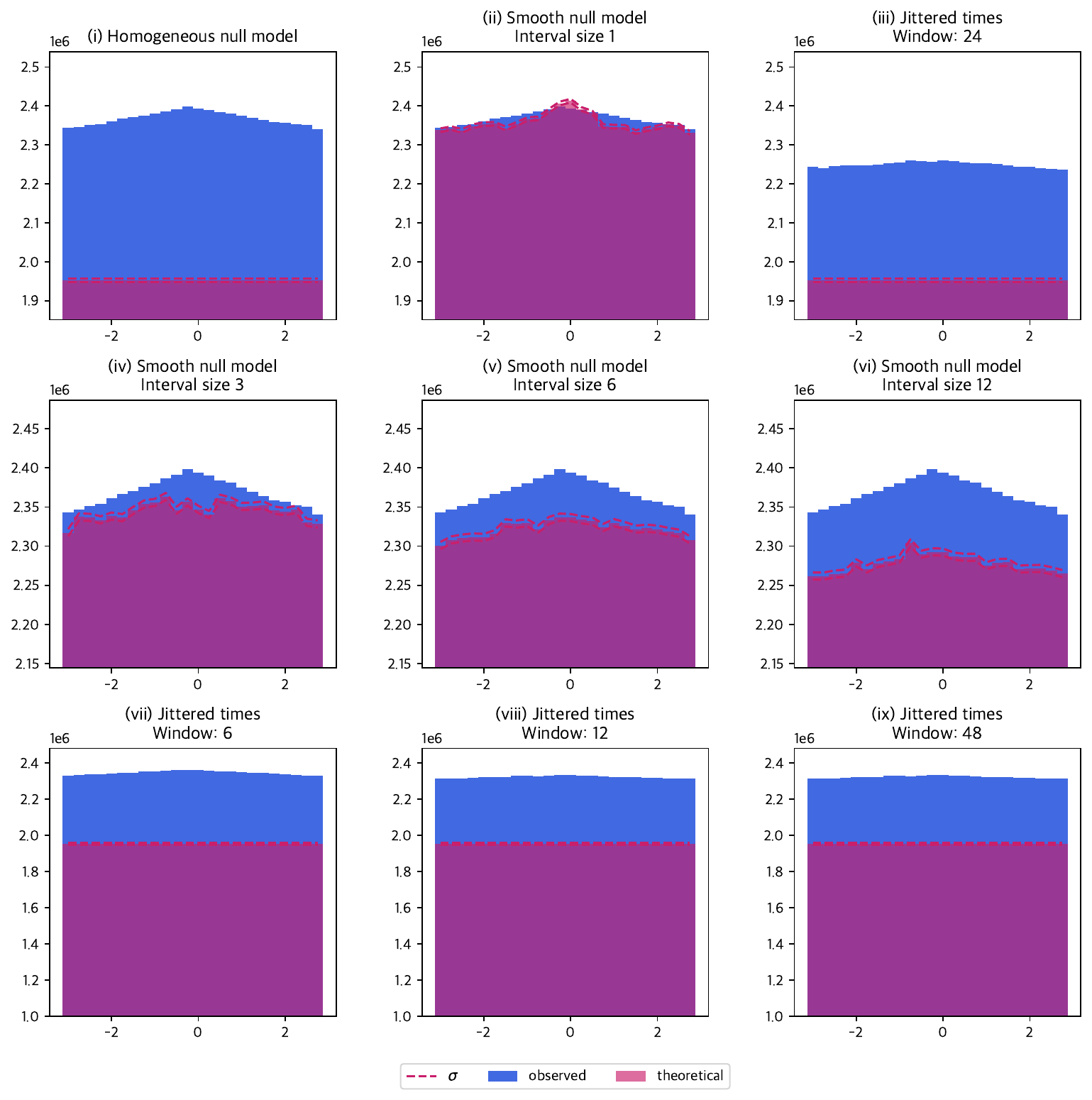}
    \caption{Bursty data is simulated using the \textit{burst ratio} model, with a \textit{burst ratio} of 0.8, a non-burst intensity of 100 with burst events occurring $U(0.05, 0.083)$ after the initial event. (i) shows the unmodified cross-correlogram with a homogeneous Poisson null model. The bursty behaviour leads to periods where the local average intensity differs greatly from the global average, leading to over or underestimates of bin heights using this approach. (ii), (iv), (v) and (vi) all show the unmodified cross-correlogram with theoretical bin heights derived using an interval jitter based intensity function. As the interval size increases, this model fits the behaviour less. In this case, the length of intervals containing bursty events is stochastic but tends to be small, meaning that smaller interval sizes better approximate the true intensity function. (iii), (vii), (viii) and (ix) modify the data by applying an interval jitter to the time series. Similarly to the smooth null model, as the jitter window increases, the intensity function of the modified data becomes more smooth, making a homogeneous null model more appropriate.}
    \label{fig:compjitter}
\end{figure}

The results discussed in the \autoref{sec:validation} show that when temporal structure is known, incorporating the correct functional form into the null model improves detection accuracy and recovery of the underlying dependency profile. Model selection, including the choice of null model, preprocessing, and spike amplitude, significantly influences performance. However, in many empirical settings the temporal structure of activity is unknown or cannot be reliably parameterized. To evaluate robustness under such conditions, we consider bursty event dynamics, a common feature of social and behavioral activity patterns \cite{de2020unraveling, sanliLocalVariationHashtag2015}.

To simulate bursty processes, we use a mechanism similar to that described in Spivak et al. \citep{spivakDeconvolutionImprovesDetection2022}. After each event, an additional event is generated with probability $BR$ (the burst ratio). If a burst event occurs, the delay is drawn from a uniform distribution on $(l, u)$. This procedure may repeat, allowing sequences of short inter-event intervals to form bursts:
\begin{equation}
    t_{i+1} = \begin{cases}
                t_i + U_i\;,\qquad\text{with probability } BR,\\
                t_i + E_i\;,\qquad\text{otherwise,}
                    \end{cases}
\end{equation}
where $U_i \sim \text{Uniform}(l, u)$ and $E_i \sim \text{Exponential}(\lambda_0)$ for baseline intensity $\lambda_0$. Under this model, burst events are recursive: the probability of the $n$-th successive burst event is $\text{BR}^n$.

To capture transitions between periods of high and low activity, the process is additionally governed by a hidden Markov model (HMM) that determines whether the system is in a bursty or non-bursty state. In the bursty state, the burst generation mechanism described above is active, whereas in the non-bursty state events follow a standard Poisson process.


Simulations follow the same procedure as before. Artificial spikes were introduced by selecting a proportion $\rho$ of source events and inserting delayed copies into the target process. Significance was assessed using the scan statistic described in \autoref{sec:validation}. Test sensitivity was evaluated across a range of spike sizes ($\rho = 0, 0.05, \ldots, 1$), with each configuration repeated over 100 runs. A visualization of the methods on a single simulation is shown in \autoref{fig:compjitter}.

\begin{figure}[t]
    \centering
    \includegraphics[width=0.48\textwidth]{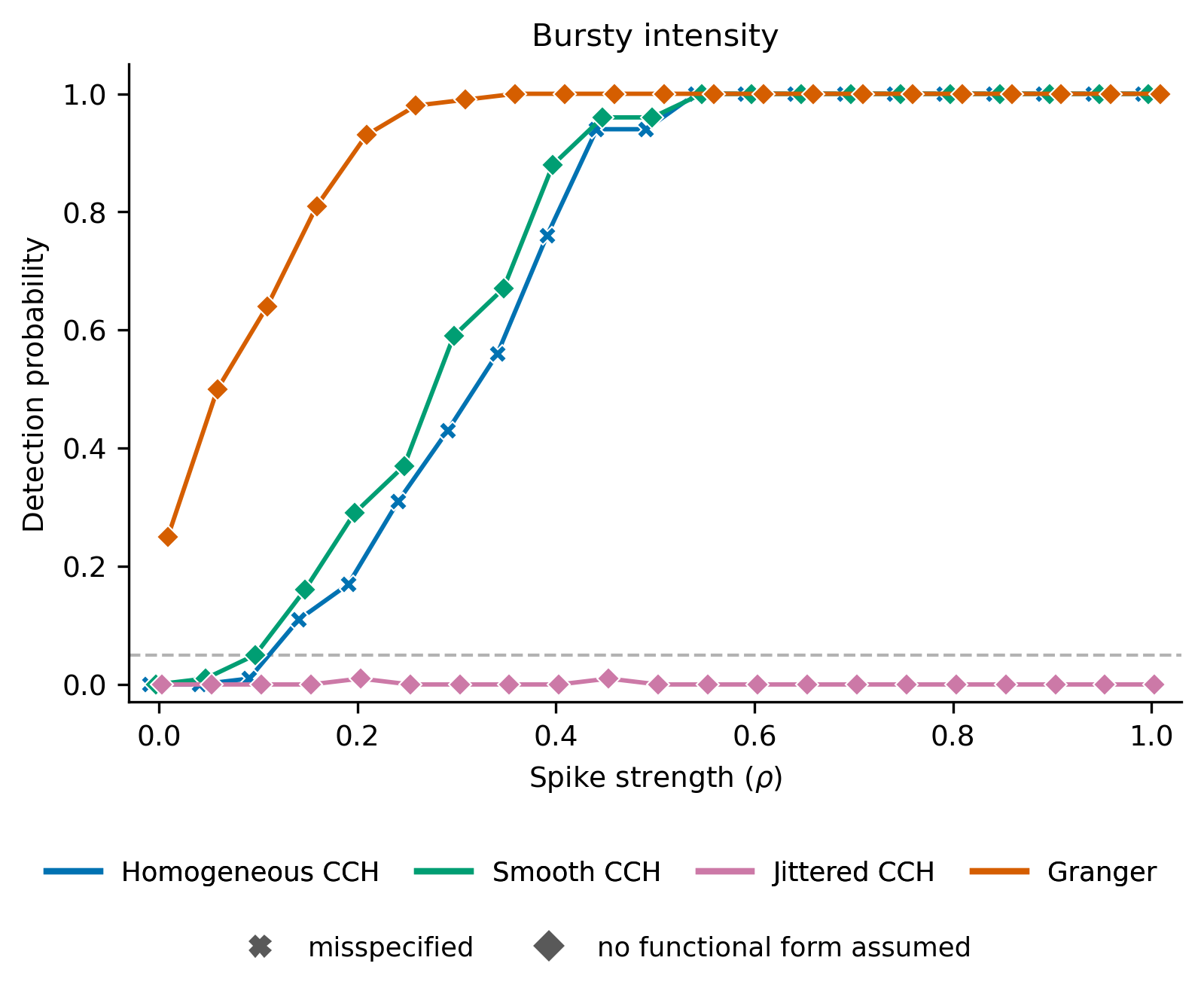}
    \caption{Detection performance as a function of interaction strength $\rho$ for simulated point processes which exhibit burstiness. This data setting was chosen to illustrate how the cross-correlogram method performs when the functional form of the data generating process is unknown or intractable. Each curve shows the proportion of simulations in which a significant temporal interaction is detected under different null models or detection methods. Bursty dynamics introduce strong local variation in event rates, which leads to systematic bias under a homogeneous Poisson null and can produce elevated false positive rates when no interaction is present ($\rho \approx 0$). Granger Causality performs well but has a high rate of false positives for $\rho=0$. Smooth inhomogeneous null models, which estimate the underlying intensity using interval-based averaging, provide improved robustness by accounting for local rate fluctuations. 
}
    \label{fig:bursty_res}
\end{figure}

The proportion of tests detecting a significant spike for each method is reported in \autoref{tab:results-bursty} and shown as a response curve in \autoref{fig:bursty_res}. As in the homogeneous and periodic settings, Granger causality does not perform well. The smooth null model and jittered cross-correlograms exhibit elevated false positive rates but true positive rates that increase with $\rho$, indicating that these methods improve detection of true temporal relationships in noisy, unspecified settings.

\begin{table}[ht]
\caption{Detection probabilities (proportion of simulations with $p<0.05$) as a function of induced spike strength $\rho$ under bursty dynamics. The smoothing or jittering interval is listed in the corresponding column.}
\label{tab:results-bursty}
\centering
\footnotesize
\setlength{\tabcolsep}{3pt}
\begin{tabular}{@{}l | cccc@{}}
\toprule
 & \multicolumn{4}{c}{Bursty data} \\
\cmidrule(lr){2-5}
$\rho$ & Const & Smooth & Jittered & Granger \\
 & & (8 hour) & (8 hour) & \\
\midrule
0.0 & 0.0 & 0.0 & 0.0 & 0.25 \\
0.05 & 0.0 & 0.01 & 0.0 & 0.5 \\
0.1 & 0.01 & 0.05 & 0.0 & 0.64 \\
0.15 & 0.11 & 0.16 & 0.0 & 0.81 \\
0.2 & 0.17 & 0.29 & 0.01 & 0.93 \\
0.25 & 0.31 & 0.37 & 0.0 & 0.98 \\
0.3 & 0.43 & 0.59 & 0.0 & 0.99 \\
0.35 & 0.56 & 0.67 & 0.0 & 1.0 \\
0.4 & 0.76 & 0.88 & 0.0 & 1.0 \\
0.45 & 0.94 & 0.96 & 0.01 & 1.0 \\
0.5 & 0.94 & 0.96 & 0.0 & 1.0 \\
0.55 & 1.0 & 1.0 & 0.0 & 1.0 \\
0.6 & 1.0 & 1.0 & 0.0 & 1.0 \\
0.65 & 1.0 & 1.0 & 0.0 & 1.0 \\
0.7 & 1.0 & 1.0 & 0.0 & 1.0 \\
0.75 & 1.0 & 1.0 & 0.0 & 1.0 \\
0.8 & 1.0 & 1.0 & 0.0 & 1.0 \\
0.85 & 1.0 & 1.0 & 0.0 & 1.0 \\
0.9 & 1.0 & 1.0 & 0.0 & 1.0 \\
0.95 & 1.0 & 1.0 & 0.0 & 1.0 \\
1.0 & 1.0 & 1.0 & 0.0 & 1.0 \\
\bottomrule
\end{tabular}
\end{table}

In bursty datasets (\autoref{fig:compjitter}), both interval jitter and smooth intensity approaches exhibit sensitivity to interval size, highlighting the advantage of explicitly modeling the functional form of temporal effects.

\section{Distribution of cross-correlogram bin heights in the non-stationary case}\label{app:binheights}

We derive the expectation and variance of bin height in the cross-correlogram for Poisson processes with sufficiently smooth intensity functions. Let $X$ and $Y$ be independent inhomogeneous Poisson processes with intensity functions $\lambda_X(t)$ and $\lambda_Y(t)$. 


For a bin size $\Delta$ consider the expected contribution to bin $i$ from a source event occuring at time $s$. The contribution to a single bin from this event is equal to the number of events which occur in target process $Y$ within $[b_i,b_i+ \Delta)$ of $s$. 

Since process $Y$ is an inhomogeneous Poisson process, the number of events in $Y$ occurring in the interval $[s+b_i,s+b_i+ \Delta)$ is Poisson distributed with mean 
$$
\Lambda_Y[s+b_i, s+b_i + \Delta) = \int_{s+b_i}^{s+b_i + \Delta} \lambda_Y (t) dt.
$$

Since events occur independently, we need to integrate this over all possible events in process $X$. This gives the expected height of bin $i$ is,
\begin{align*}
\mathbb{E}[h_i] = \int_w^{T-w} \lambda_X(s) \int_{s+b_i}^{s+b_i + \Delta} \lambda_Y (t) dt ds 
\end{align*}

This is a continuous measure, but taking the average over the bin interval, $[b_i,b_i+\Delta)$ gives a discrete theoretical value. Conditioned on event times in $X$, the conditional variance is equal to the conditional mean. 

Conditioned on the set of $E_X$ event times the contributions to $h_i$ are independent Poisson variables, hence
\[
\text{Var}(h_i\mid E_X)=\mathbb{E}[h_i\mid E_X].
\]
Unconditionally, by the law of total variance,
\[
\text{Var}(h_i)=\mathbb{E}\big[\text{Var}(h_i\mid E_X)\big]+\text{Var}\big(\mathbb{E}[h_i\mid E_X]\big),
\]
The second term is generally nonzero under inhomogeneous intensities; it vanishes in the homogeneous case, yielding \(\text{Var}(h_i)=\mathbb{E}[h_i]\).








\section{Expected difference between jittered and original spike trains}\label{app:jittererror}

Consider the difference between $\int_{s_i+r_i}^{s_i+r_i + \Delta} \lambda_T (t) dt$ and $ \int_{s_i+r_i}^{s_i+r_i + \Delta} \Tilde{\lambda}_t(t) dt$.

\begin{equation}
\label{eqn:errorlambda}
e(s_i+r_i) = \int_{s_i+r_i}^{s_i+r_i + \Delta} \lambda_T (t) dt - \int_{s_i+r_i}^{s_i+r_i + \Delta} \Tilde{\lambda}_t(t) dt 
\end{equation}

We can make the simplifying assumption that $\delta>\Delta$, i.e. that the jitter interval is larger than the width of the cross-correlogram bins. This is a reasonable assumption as for $\delta$ much smaller than $\Delta$, the difference between the expected height of the cross correlogram bins is negligible as multiple intervals will occur in a target time interval of width $\Delta$ and the error will only be due to jitter intervals which overlap the edges of the cross-correlogram target bin.

\begin{figure}[H]
    \centering
    \includegraphics[width=0.4\textwidth]{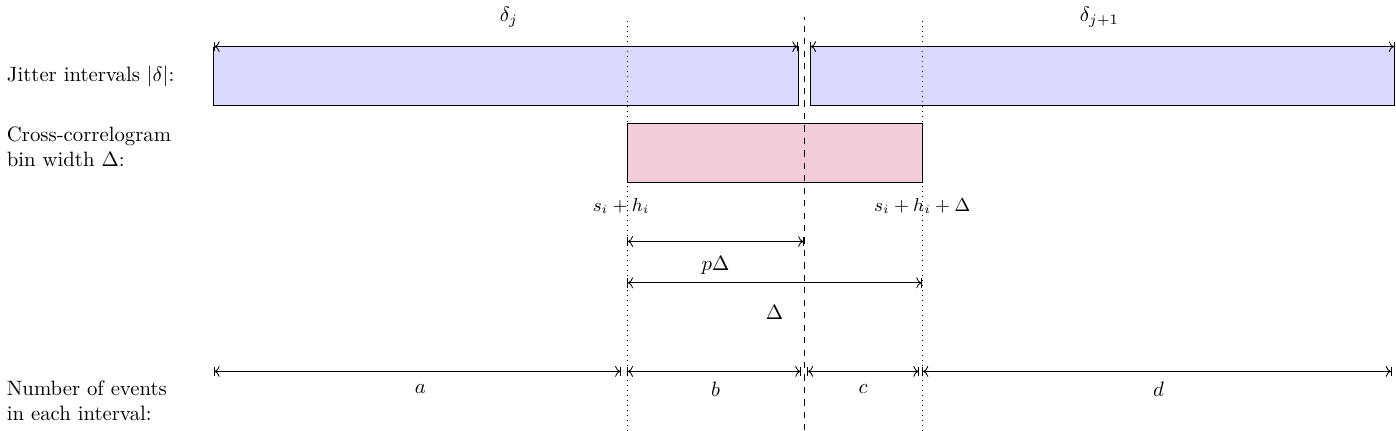}
    \caption{Setup for considering the difference between original and jittered cross correlograms.}
    \label{fig:jitterdifferencesetup}
\end{figure}

The error introduced by jitter is due to the regions where the jitter bin falls outside the cross-correlogram target bin. Let the expected number of events assigned to the cross-correlogram target bin from the start of the left jitter-interval to the start of the target time interval be given by $a$, and the expected number from the end of the target time interval to the end of the right-most jitter interval be given by $d$. $a$ and $d$ represent the events which are included in $\Tilde{\lambda}_t(t)$ but not in $\lambda(t)$. The events in the target time interval will be divided up into two sets, $b$ and $c$ which fall into the two jitter-intervals. This setup is visually represented in \autoref{fig:jitterdifferencesetup}.

The total number of events in the interval $[s_i+r_i,s_i+r_i+\Delta]$,  is given by $\int_{s_i+r_i}^{s_i+r_i + \Delta} \lambda_T (t) dt = b+c$.

We also have that 
$$
\Tilde{\lambda}_j(t) = \dfrac{a+b}{\delta}
$$
$$
\Tilde{\lambda}_{j+1}(t) = \dfrac{c+d}{\delta}
$$

Hence, 
\begin{align*}
    \int_{s_i+r_i}^{s_i+r_i + \Delta} \widetilde{\lambda}_t(t)\,dt
    &=
    \begin{aligned}[t]
        &\int_{s_i+r_i}^{s_i+r_i + p\Delta} \widetilde{\lambda}_j(t)\,dt \\
        &\quad + \int_{s_i+r_i+p\Delta}^{s_i+r_i+\Delta}
        \widetilde{\lambda}_{j+1}(t)\,dt
    \end{aligned} \\
    &= p\Delta \dfrac{a+b}{\delta}
    + (1-p)\Delta \dfrac{c+d}{\delta}.
\end{align*}

Substituting into \autoref{eqn:errorlambda}, for $\delta = k \Delta$,
$$
e(s_i+r_i) = -\dfrac{p}{k} (a+b) - \dfrac{1-p}{k} (d-c)
$$




In cases where the intensity function is approximately constant, i.e. $a, d \approx k \; O(b,c).$, the error is small. This is the case in a stationary process. In general, as $a, d$ get more different than $b$ and $c$, the approximation given by the jittered version gets worse and error increases.

\subsection{Cross-correlogram error}

Using a simplified set up, we can quantify the size of the error introduced by interval jitter when compared to a homogeneous Poisson null model when periodic non-stationary or bursty effects are present in the data. 

We will evaluate the size of the error introduced interval jittering of the target process by looking at the theoretical difference between bin heights for the cross-correlogram constructed using the original and jittered time series.


If the original height of bin $j$ is given by $h_{j}$, let the height from the jittered process be denoted $\tilde{h}_{j}$. To produce a meaningful measure of error and minimize the effects of scale due to parameter choice, we will consider the relative error introduced by the jitter procedure. 

$$
e_{j} = \dfrac{| h_{j}-\tilde{h}_{j}|}{h_{j}}
$$

As the distribution of the errors is not known, we will consider the individual errors generated by each bin as well as the median error. An `error band' will be drawn by considering the $5^\text{th}$ and $95^\text{th}$ percentiles of the individual bin errors. 

A value of $e_{j}=0.01$ represents a 1\% difference between the final height of the original and jittered cross-correlograms for a single bin. Even small changes in error can cause the power of statistical tests to reduce greatly, so the impacts of error propagating through the cross-correlogram procedure should be considered.


For periodic non-stationary effects, the theoretical bin heights are known, so the errors can be derived exactly for each bin. For bursty effects, simulations will be used to approximate the error for each bin and cross-correlogram.

\subsection{Periodic non-stationary effects}

Let the source intensity function be a constant Poisson intensity, as given in \autoref{eq:constint}.

\begin{equation}\label{eq:constint}
     \lambda_S(t) = c_S
\end{equation}

We will represent periodic non-stationary target intensity function using a sinusoidal function with amplitude $a_T$ and period $b_T$. The constant background intensity is 1, so a value of $a_T=2$ can be interpreted as an additional periodic intensity with amplitude of twice the background intensity.

\begin{equation}\label{eq:periodicns}
     \lambda^\star_T(t) = c_T + a_T \sin \left(\dfrac{2 \pi  t}{ b_T}\right)
\end{equation}

\begin{figure*}
    \centering
    \includegraphics[width=\textwidth]{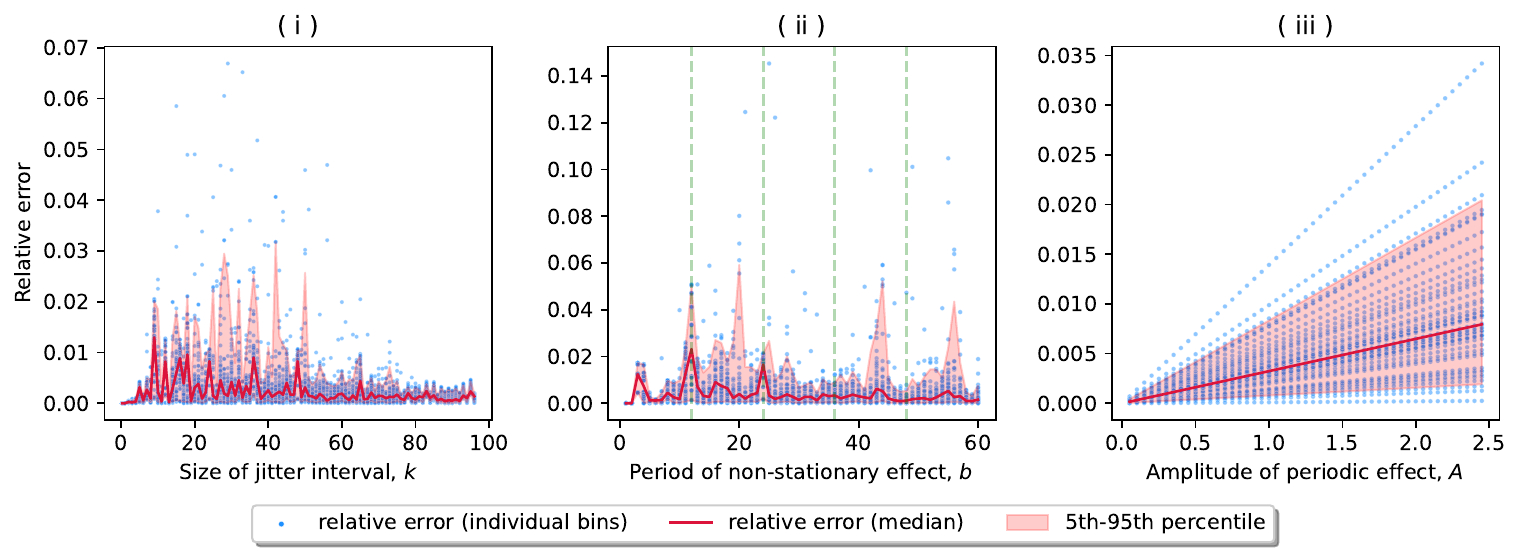}
    \caption{(i) Relative Error as a function of the size of the jitter window, $k$. (ii) Relative Error as a function of amplitude of periodic non-stationary effects, $A$. (iii) Relative Error as a function of period  of periodic non-stationary effects, $b$.} 
    \label{fig:jitter_periodic}
\end{figure*}

We can see how the relative error, $e_{j}$ changes with each of these features (\autoref{fig:jitter_periodic}). For each of these simulations, the following parameters are used as default values: $c_S$ $=$ $1$, $c_T$ $=$ $1$, $T$ $=$ $1000$, $\Delta = 0.25$, $a_T$ $=$ $1$, $b_T$ $=$ $24$, $w=6$, $k=24$. i.e. a cross-correlogram which has a width of 6 units, on average we see one event per unit, the bin widths are 0.25 units, the period is 24 units and the amplitude of the periodic non-stationarity is 1. The default jitter interval is 24 units.


With this setup, we find that:
\begin{enumerate}
    \item There does not seem to be a strong relationship between error and the size of the jitter interval. There is a slight trend that error increases as the size of the jitter window increases. The maximum individual error increases with jitter interval size. \autoref{fig:jitter_periodic} (i).
    \item Both the median and spread of the individual errors grow with increasing amplitude of periodic non-stationary effects. \autoref{fig:jitter_periodic} (ii).
    \item There appears to be a relationship between error and period, with stronger effects at some periods (e.g. 12 units, 20 units). \autoref{fig:jitter_periodic} (iii).
\end{enumerate}

\begin{figure*}
    \centering
    \includegraphics[width=\textwidth]{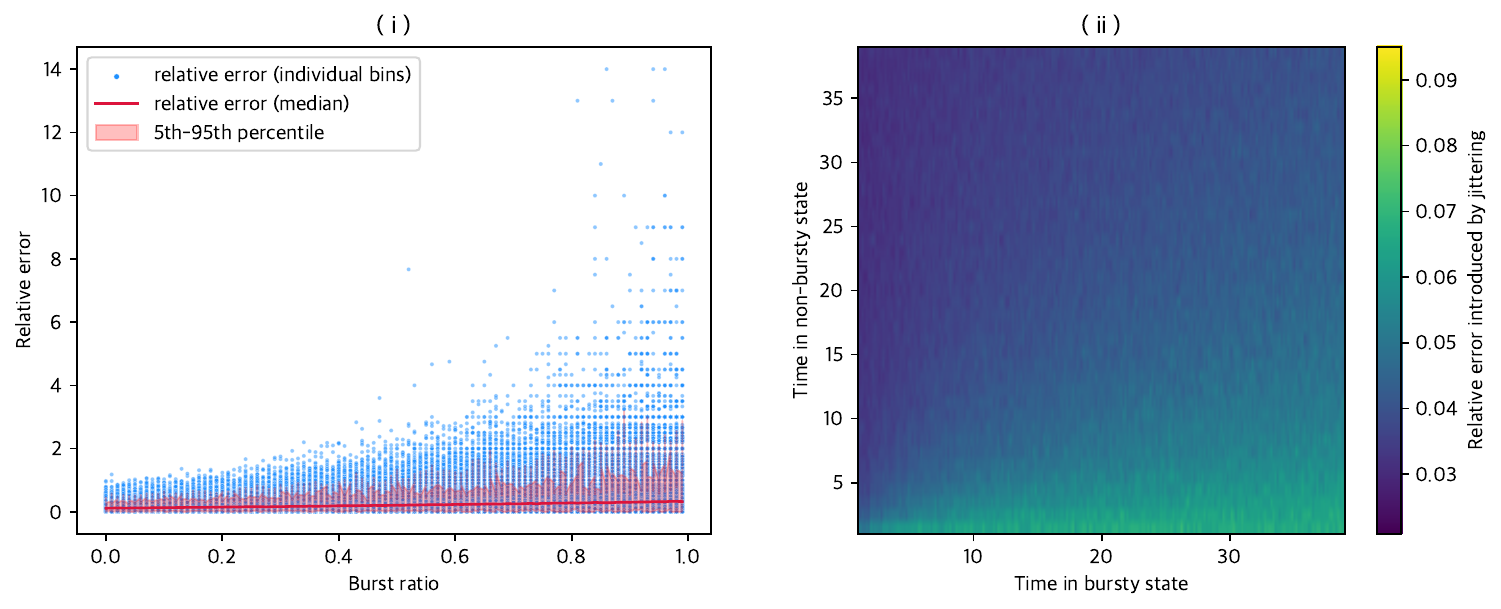}
    \caption{For the burst ratio and HMM bursty event models, the difference in bin height for the original and jittered process has been simulated. (i) shows results from the burst ratio model, demonstrating that the difference grows as the burst ratio increases. (ii) shows the HMM model. As time spend in a bursty state increases, the difference introduced by jittering grows.} 
    \label{fig:jitter_br_error}
\end{figure*}

In \autoref{fig:jitter_br_error} we consider both models for bursty effects (burst ratio and HMM), showing that as burstiness increases, the bias introduced by jittering grows.

\section{Topic classifications}\label{app:topicclassifications}

To categorize hashtags collected from Twitter, we used ChatGPT (OpenAI, GPT-4o \citep{openai_chatgpt_2023}) to assign each hashtag a general topic and a more specific subtopic. Following this, these classifications were manually checked and some topics were merged to ensure the topics were consistent and interpretable. The prompt which was used is given below.

\texttt{Can you assign a topic and subtopic to each word in a list for me? These words represent hashtags from the social media platform X / Twitter. The topics you assign should be general (for example: music, sport, politics, news, entertainment, \\technology, health, finance, environment, social media,\\ literature), and then the subtopics should be more specific (for example, 'bts' should have the topic 'music' and the\\ subtopic 'K-pop').}

\texttt{I will use these topic labels to analyse the dynamics of these hashtags as part of thematic groups.}

\texttt{Please return the output in the format:
[(hashtag, topic, subtopic), ...]}

\texttt{For example:
[('bts', 'music', 'K-pop'), ('f1', 'sport', 'Formula 1'), ('trump', 'politics', 'US politics')]}

\texttt{Do you have any questions before I give you the list of hashtags to ensure the topic labels are consistent and meaningful?}

The final classifications are given below.
\small
\bgroup
\def\arraystretch{1.5}
\begin{center}
\begin{tabular}{ p{0.15\linewidth}  p{0.15\linewidth}   p{0.55\linewidth}}
Topic & Unique hashtags & Hashtags \\
music & 29 &
\#ateez, \#bestcoversong, \#blackpink, \#boywithluv, \#bts,
\#btsarmy, \#edm, \#exo, \#listen, \#monsta\_x, \#monstax,
\#music, \#nct, \#nowplaying, \#redvelvet, \#seventeen,
\#soundcloud, \#straykids, \#twice, \#갓세븐, \#몬스타엑스,
\#방탄소년단, \#블랙핑크, \#세븐틴, \#스트레이키즈,
\#에이티즈, \#트와이스, \#got7, \#nct127 \\
politics & 23 &
\#assange, \#auspol, \#ausvotes, \#brexit, \#cdnpoli,
\#china, \#hongkong, \#iran, \#istandwithdan, \#lnp,
\#maga, \#morrison, \#nswpol, \#pell, \#qldpol,
\#robodebt, \#scomo, \#scottyfrommarketing, \#sportsrorts,
\#springst, \#trump, \#watergate, \#ausvotes2019 \\
celebrities & 18 &
\#bambam, \#jhope, \#jimin, \#jin, \#jk, \#jungkook,
\#peckpalitchoke, \#perthtanapon, \#rm, \#suga, \#taehyung, \#v,
\#\textthai{เป๊กผลิตโชค}, \#\textthai{รอยยิ้มของชูครีม}, \#\textthai{ออฟกัน},
\#정국, \#지민, \#태형 \\
news and current affairs & 11 &
\#breaking, \#fakenews, \#insiders, \#news, \#qanda,
\#rubyprincess, \#thedrum, \#4corners, \#7news, \#9news,
\#abc730 \\
television & 10 &
\#gameofthrones, \#mafs, \#mafsau, \#masterchefau,
\#neighbours, \#sanditon, \#savesanditon, \#shadowhunters,
\#survivorau, \#thebachelorau \\
general & 10 &
\#gay, \#job, \#love, \#onthisday, \#quote, \#raw,
\#socialmedia, \#spub, \#stilltheone, \#stop \\
literature & 10 &
\#amreading, \#amwriting, \#asmsg, \#book, \#books,
\#iartg, \#ibooks, \#romance, \#writingcommunity, \#ian1 \\
gaming & 9 &
\#acnh, \#animalcrossing, \#fgo, \#fortnite, \#gamedev,
\#nintendoswitch, \#twitch, \#ps4live, \#ps4share \\
health & 8 &
\#coronavirus, \#covid, \#health, \#mentalhealth,
\#covid\_19, \#covid--19, \#covid19, \#covid19aus \\
music awards & 8 &
\#bbmastopsocial, \#bestfanarmy, \#grammys, \#iheartawards,
\#mamavote, \#mtvhottest, \#pcas, \#twitterbestfandom \\
adult & 8 &
\#boobs, \#findom, \#horny, \#milf, \#nsfw, \#porn, \#sex, \#sexy \\
\end{tabular} 
\end{center}
\egroup
\bgroup
\def\arraystretch{1.5}
\begin{center}
\begin{tabular}{ p{0.15\linewidth}  p{0.15\linewidth}   p{0.55\linewidth}}
Topic & Unique hashtags & Hashtags\\
sports & 8 &
\#afl, \#ashes, \#ausopen, \#mufc, \#vasc, \#ao2020,
\#nrl, \#f1 \\
travel & 8 &
\#adelaide, \#australia, \#brisbane, \#canberra,
\#melbourne, \#perth, \#sydney, \#travel \\
environment & 6 &
\#bushfires, \#climate, \#climatechange, \#climateemergency,
\#nature, \#nswfires \\
finance & 5 &
\#asx, \#ausbiz, \#bitcoin, \#btc, \#crypto \\
other & 28 & \#avengersendgame, \#blog, \#criticalrole, \#podcast, 
\#free, \#giveaway, \#marketing, \#seo, 
\#follow, \#newprofilepic, \#qt, \#rt, 
\#art, \#kaye\_menner, \#photography, \#obamagate, 
\#qanon, \#wwg1wga, \#blacklivesmatter, \#climatestrike, 
\#soompiawards, \#teenchoice, \#business, \#leadership,
\#ai, \#kindle, \#weather, \#bob19vehs
\end{tabular} 
\end{center}
\egroup
 
\end{appendices}

\end{document}